\documentclass[12pt,hidelinks]{article}
\usepackage{mathrsfs,amsmath,epsfig,amssymb,color,graphicx,bm}
\usepackage{algorithm,algorithmic,natbib,subcaption,hyperref,float}
\usepackage[margin=1in]{geometry}
\usepackage[symbol]{footmisc}
\usepackage{rotating}
\graphicspath{{figures/}}
\allowdisplaybreaks

\newtheorem{theorem}{Theorem}

\newcommand{\bbeta}{{\boldsymbol \beta}}

\newcommand{\bGamma}{{\boldsymbol \Gamma}}
\newcommand{\bSigma}{{\boldsymbol \Sigma}}

\newcommand{\bfeta}{{\boldsymbol \eta}}
\newcommand{\bzeta}{{\boldsymbol \zeta}}

\newcommand{\bgamma}{{\boldsymbol \gamma}}
\newcommand{\btheta}{{\boldsymbol \theta}}
\newcommand{\bphi}{{\boldsymbol \phi}}
\newcommand{\bTheta}{{\boldsymbol \Theta}}

\begin{document}

\centerline {\Large\bf A Latent Logistic Regression Model }
\centerline {\Large\bf with Graph Data }
\vspace*{0.2in}
\centerline{Haixiang Zhang$^*$, Yingjun Deng, Alan J.X. Guo, Qing-Hu Hou and Ou Wu}

 \vspace*{0.1in}

\centerline{\it \small $^{}$Center for Applied Mathematics, Tianjin University, Tianjin 300072, China}

\footnotetext[1]{Corresponding author: haixiang.zhang@tju.edu.cn (Haixiang Zhang)}
\vspace{0.5cm}

\begin{abstract}

Recently, graph (network) data is an emerging research area in artificial intelligence, machine learning and statistics. In this work, we are interested in whether node's labels (people's responses) are affected by their neighbor's features (friends' characteristics).  We propose a novel latent  logistic regression model to describe the network dependence with binary responses. The key advantage of our proposed model is that a latent binary indicator is
introduced to indicate  whether a node is susceptible to the influence of its neighbour. A score-type test
is proposed to diagnose the existence of network dependence. In addition,  an EM-type algorithm is used to estimate the model parameters under network dependence.  Extensive simulations are conducted to evaluate the performance of our method. Two public datasets are used to illustrate the effectiveness of the proposed latent logistic regression model.\\

\end{abstract}

\section{Introduction}

Nowadays, graphs or networks are widely used in many research fields, such as social interactions, protein-protein interactions, chemical molecule bonds, transport networks, etc.  Great efforts have been focused on network data in the literature. For example, \cite{Zhu-AOS-2017}  proposed a network vector autoregressive  model.   \cite{YanT-JASA-2019}  studied the maximum likelihood estimation of a directed network model with covariates. \cite{Zhang-JCGS-2022} studied some topics on large scale social networks. \cite{Chandna-2021} considered the local linear estimation of the graphon function. \cite{Zhu-JASA-2021} introduced a network functional varying coefficient model. \cite{Pan-SII-2022} proposed  a latent space logistic regression model for link prediction with  social networks.  \cite{Zhao-Biom-2022} proposed a dimension reduction method for covariates in network data, etc.

The logistic regression is a very famous statistical tool, which plays an important role in practical applications. e.g. finance research \cite[]{Yazhe-2019}, public health \cite[]{Stephenie-2003}, medicine \cite[]{Minggen-JDS-2012},
education \cite[]{Sandra-1993}, bioinformatics \cite[]{WuTT-2009}. There have been some methodological developments on the logistic regression. e.g. \cite{Landwehr-JASA-1984} proposed a graphical methods for assessing logistic regression models.
\cite{Stefanski-AOS-1985}  studied the logistic regression model when covariates are subject to measurement error.
\cite{Jennings-1986-JASA}  investigated the outliers and residual distributions in logistic regression. \cite{Efron-JASA-1988}
used the logistic regression techniques to estimate hazard rates and survival curves from censored data. \cite{Carroll-JRSSB-1993}  investigated robustness in the logistic regression model. \cite{Meier-JRSSB-2008} extended the group lasso to logistic regression models. \cite{jun-2009} proposed an algorithm for solving large-scale sparse logistic regression.  \cite{SIAM-C-L-2009} examined the problem of efficient feature  selection for logistic regression on very large data sets. \cite{Shi-JMLR-2010} and \cite{Yuan-JMLR-2012} studied some algorithms for L1-regularized logistic regression.  \cite{Bryan-PMLR-2012} proposed a fast and exact model selection procedure for L2-regularized logistic regression. \cite{AI-S_Das-2013}  introduced  new supervised and semi-supervised learning algorithms based on locally-weighted logistic regression.  \cite{ADS-Yogesh-2017} considered estimation of the probability density function and the cumulative distribution function of the generalized logistic distribution. \cite{Andrew-2007} and \cite{Pattern-2018} studied the active learning methods for logistic regression. \cite{wang-ICML-2020} studied binary logistic regression for rare events data.
\cite{Kyoohyung-22019} presented an efficient algorithm for logistic regression on homomorphic encrypted data.
\cite{wang2018optimal} and \cite{zuo2021optimal} considered the optimal subsmapling for logistic regression in big data, among others.

 However, the above-mentioned logistic regression methods did not consider the effects of networks. For example,  does a persion like to  playing a game given that his/her friends
like it?  Will the customers by  a commodity if their friends have  bought it?
In this work, we are interested in exploring whether certain type of network dependence exists
with binary outcomes data and to quantify this dependence structure if it exists.  To deal with this issue,
we propose a  logistic model with latent binary indicator, which has the ability to describe whether a node is susceptible to the influence of its neighbor.  Our method has the following two advantages:
First, the proposed logistic model with a latent binary indicator is very flexibility in practical
applications.  It provides a solution to estimate the probability that a node might be
affected by neighbor's characteristics in the network. Hence, we can detect a subgroup of nodes who are more likely to be influenced by their neighbors.  Second, we give a score-type test for detecting the existence of the network
dependence in the logistic model. An EM algorithm is employed to estimate the model parameters, which leads to consistence estimator with desirable performance in simulations.

The remainder of this article is organized as follows: In Section 2, we introduce a latent logistic regression model. In Section 3, we propose a  supremum score test statistic to detect the existence of network dependence. In Section 4,  an EM-type estimation algorithm is proposed.  Simulations and a real data application are presented in Sections 5 and 6, respectively. In Section 7, we give some concluding remarks.

\section{Model and Notation}

Let $G = (V, E)$ be an undirected graph with nodes $V$ and edges $E$.  Assume that there are $n$
nodes belonging to two classes.  Let $Y_i \in \{0,1\}$ and $\mathbf{X}_i= (X_{i1},\cdots,X_{ip})^\prime \in \mathbb{R}^p $ be the binary label and feature vector of the $i$-th node $v_i$, $i=1,\cdots,n$. Given the graph structure of $G$, we propose a novel
latent logistic regression model:
\begin{align}\label{M-1}
{\mathbb{P}}(Y_i = 1|\mathbf{X}_i, \zeta_i) &= \frac{\exp\{\beta_0 + \mathbf{X}_i^\prime \bbeta + \delta \zeta_i \sum_{j=1}^n a_{ij}\mathbf{X}_j^\prime \bbeta\}}{ 1+ \exp\{\beta_0 + \mathbf{X}_i^\prime \bbeta + \delta \zeta_i \sum_{j=1}^n a_{ij}\mathbf{X}_j^\prime \bbeta\}},~~ i=1,\cdots,n,
\end{align}
where $\beta_0 \in \mathbb{R}$ is an intercept, $\bbeta = (\beta_1,\cdots,\beta_p)^\prime$ is the vector of regression parameters, $\mathbf{A} = (a_{ij})$ is the adjacency matrix ($a_{ii} =0$, and $a_{ij} = 1$ if there is an edge between the $i$th node and $j$th node, $a_{ij} =0$ otherwise); $\zeta_i \in \{0,1\}$ is a latent indicator
denoting whether the label (response) of $i$th node depends on its neighbor's features. Note that $\zeta_i$ is unobservable, and we assume that
\begin{align}\label{M-2}
{\mathbb{P}}(\zeta_i = 1|\mathbf{X}_i)= \frac{\exp\{\gamma_0 + \mathbf{X}_i^\prime \bgamma\}}{1+ \exp\{\gamma_0 + \mathbf{X}_i^\prime \bgamma\}}.
\end{align}

The parameter $\delta$ plays the role of describing the magnitude of the dependence of a node to its neighbor.
When $\delta = 0$, there is no  network dependence between labels of
connected nodes, and the parameters $\gamma_0$ and $\bgamma$ are not estimable in this case. In what follows, we
will proposed a method to test the  null hypothesis $H_0: \delta =0$, and then give an
EM-type algorithm to estimate the model parameters under $\delta \neq0$.

\section{Test for $H_0: \delta =0$}
\setcounter{equation}{0}
First we assume that $\boldsymbol{\zeta}= (\zeta_1, \cdots,\zeta_n)^\prime$ is known. The log likelihood function is
\begin{eqnarray}\label{LF}
L(\btheta;\bzeta) &=& \sum_{i=1}^n\Bigg[Y_i\Big(\beta_0 + \mathbf{X}_i^\prime \bbeta + \delta \zeta_i \sum_{j=1}^n a_{ij}\mathbf{X}_j^\prime \bbeta\Big)\nonumber\\
 &&- \log\bigg(1+ \exp\Big\{\beta_0 + \mathbf{X}_i^\prime \bbeta + \delta \zeta_i \sum_{j=1}^n a_{ij}\mathbf{X}_j^\prime \bbeta\Big\} \bigg)\Bigg],
\end{eqnarray}
where $\btheta = (\delta, \beta_0, \bbeta^\prime)^\prime$. Under $H_0$, model (\ref{M-1}) reduces to the standard logistic model. Let $\tilde{\beta}_0$ and $\tilde{\bbeta}$ denote the maximum likelihood estimator under the
null.  Denote $\tilde{\btheta} = (0,\tilde{\beta}_0,\tilde{\bbeta}^\prime)^\prime$.  Some calculations lead to the following score function:
\begin{eqnarray}\label{Eq-3}
S(\tilde{\btheta};\bzeta) = \frac{\partial L(\btheta;\bzeta)}{\partial \delta}|_{\btheta =\tilde{\btheta}} = \sum_{i=1}^n \tilde{Z}_i\left[Y_i - \frac{ \exp(\tilde{\beta}_0 + \mathbf{X}_i^\prime \tilde{\bbeta})}{1+\exp(\tilde{\beta}_0 + \mathbf{X}_i^\prime \tilde{\bbeta})}\right],
\end{eqnarray}
where $\tilde{Z}_i = \zeta_i \sum_{j=1}^n a_{ij}\mathbf{X}_j^\prime \tilde{\bbeta}$.  Because $\bzeta$ is not available in $\tilde{Z}_i$'s, we propose to replace $\zeta_i$ with its expectation $E(\zeta_i) = P(\zeta_i = 1|\mathbf{X}_i)$ given in (\ref{M-2}). For convenience, denote $\tilde{Z}_i^* = E(\zeta_i) \sum_{j=1}^n a_{ij}\mathbf{X}_j^\prime \tilde{\bbeta}$. After replacing $\tilde{Z}_i$ with $\tilde{Z}_i^*$ in (\ref{Eq-3}),
we obtain a score-type statistic:
\begin{eqnarray}\label{Eq-4}
S^*(\tilde{\btheta};\bphi) = \sum_{i=1}^n \tilde{Z}^*_i\left[Y_i - \frac{ \exp(\tilde{\beta}_0 + \mathbf{X}_i^\prime \tilde{\bbeta})}{1+\exp(\tilde{\beta}_0 + \mathbf{X}_i^\prime \tilde{\bbeta})}\right],
\end{eqnarray}
where $\bphi=(\gamma_0,\bgamma^\prime)^\prime$.

Motivated by \cite{FanSL-JASA-2017}, we propose a
supremum score test statistic:
\begin{eqnarray}
T_n = \sup\limits_{\bphi \in \bGamma} \frac{\{S^*(\tilde{\btheta};\bphi)\}^2}{\sum_{i=1}^n \{U_i^*(\tilde{\btheta};\bphi)\}^2},
\end{eqnarray}
where $\bGamma\in \mathbb{R}^{p+1}$, and
\begin{eqnarray*}
U_i^*(\tilde{\btheta};\bphi) &=&  \left[Y_i - \frac{ \exp(\tilde{\beta}_0 + \mathbf{X}_i^\prime \tilde{\bbeta})}{1+\exp(\tilde{\beta}_0 + \mathbf{X}_i^\prime \tilde{\bbeta})}\right] \left\{\tilde{Z}^*_i
 - \mathcal{B}_n^*(\tilde{\btheta})\mathcal{I}_n^{-1}(\tilde{\btheta})(1,\mathbf{X}_i^\prime)^\prime\right\},
\end{eqnarray*}
where  $\mathcal{I}_n (\tilde{\btheta})= -\frac{\partial^2 L(\btheta;\bzeta)}{\partial \bfeta \bfeta^\prime}|_{\btheta=\tilde{\btheta}}$, $\mathcal{B}_n(\tilde{\btheta})= -\frac{\partial^2 L(\btheta;\bzeta)}{\partial \delta \partial\bfeta^\prime}|_{\btheta=\tilde{\btheta}}$, $\bfeta  = (\beta_0, \bbeta^\prime)^\prime$  and $\mathcal{B}^*_n(\tilde{\btheta})$ is given from $\mathcal{B}_n(\tilde{\btheta})$ by replacing  $\zeta_i$ with its expectation $E(\zeta_i) = P(\zeta_i = 1|\mathbf{X}_i)$. To be more specific, we have the following two explicit expressions:
\begin{eqnarray*}
\mathcal{I}_n (\tilde{\btheta}) =  \sum_{i=1}^n \left[\frac{\exp(\tilde{\mathbf{X}}^\prime_i\tilde{\bfeta})}{1 +\exp(\tilde{\mathbf{X}}^\prime_i\tilde{\bfeta})} - \left\{\frac{\exp(\tilde{\mathbf{X}}^\prime_i\tilde{\bfeta})}{1 +\exp(\tilde{\mathbf{X}}^\prime_i\tilde{\bfeta})}\right\}^2\right]\tilde{\mathbf{X}}_i\tilde{\mathbf{X}}^\prime_i,
\end{eqnarray*}
and
\begin{eqnarray*}
\mathcal{B}_n(\tilde{\btheta}) &=& -\sum_{i=1}^n \Bigg[\zeta_i \bigg\{ Y_i - \frac{\exp(\tilde{\mathbf{X}}^\prime_i\tilde{\bfeta})}{1 +\exp(\tilde{\mathbf{X}}^\prime_i\tilde{\bfeta})}\bigg\}\Big(0,\sum_{j=1}^n a_{ij}\mathbf{X}_j^\prime\Big)\\
&&+ \zeta_i \sum_{j=1}^n a_{ij}\mathbf{X}_j^\prime \tilde{\bbeta}\bigg( \bigg\{\frac{\exp(\tilde{\mathbf{X}}^\prime_i\tilde{\bfeta})}{1 +\exp(\tilde{\mathbf{X}}^\prime_i\tilde{\bfeta})}\bigg\}^2 -\frac{\exp(\tilde{\mathbf{X}}^\prime_i\tilde{\bfeta})}{1 +\exp(\tilde{\mathbf{X}}^\prime_i\tilde{\bfeta})} \bigg)\mathbf{\tilde{X}}_i^\prime\Bigg],
\end{eqnarray*}
where $\tilde{\bfeta}  = (\tilde{\beta}_0, \tilde{\bbeta}^\prime)^\prime$, and $\tilde{\mathbf{X}}_i = (1,\mathbf{X}_i^\prime)^\prime$.

\begin{theorem}
As $n\rightarrow \infty$, we have $T_n$ converges in distribution to $\sup\limits_{\bphi \in \bGamma} G^2(\bphi)$ under $H_0$, where $\{G(\bphi): \bphi \in \bGamma\}$ is a mean zero Gaussian process with the covariance function
\begin{eqnarray*}
\bSigma(\bphi_1,\bphi_2)=\lim_{n\rightarrow \infty}\frac{\sum_{i=1}^n U_i^*(\btheta_t;\bphi_1)U_i^*(\btheta_t;\bphi_2)}{[\sum_{i=1}^n \{U_i^*(\btheta_t;\bphi_1)\}^2\sum_{i=1}^n \{U_i^*(\btheta_t;\bphi_2)\}^2]^{1/2}},
\end{eqnarray*}
for any $\bphi_1$, $\bphi_2\in \bGamma$.
\end{theorem}

We adopt a resampling method in order to obtain the the critical value of the
asymptotic  distribution of $T_n$ under $H_0$. To be more specific, we define a perturbed test statistic:
\begin{eqnarray}\label{PTS-3.5}
T_n^* = \sup\limits_{\bphi \in \bGamma} \frac{\{\sum_{i=1}^n\xi_iU_i^*(\tilde{\btheta};\bphi)\}^2}{\sum_{i=1}^n \{U_i^*(\tilde{\btheta};\bphi)\}^2},
\end{eqnarray}
where $\xi_1,\cdots,\xi_n$ are independently generated from $N(0,1)$. Note that $T_n$ and $T_n^*$ own the same asymptotic distribution under $H_0$. By repeatedly generating a great deal of perturbed statistics, we can obtain the empirical upper $\alpha$-quantile, $C_\alpha$, of the perturbed statistics $T_n^*$'s. The null $H_0$ is rejected if $T_n > C_\alpha$.

\section{Estimation Method}
\setcounter{equation}{0}
In this section, we assume $\delta \neq 0$. Therefore, the models (\ref{M-1}) and (\ref{M-2}) are identifiable
in terms of parameter $\bTheta=(\delta, \beta_0, \bbeta^\prime, \bphi^\prime)^\prime$. The complete log likelihood function is
 \begin{eqnarray}\label{Eq-3.1}
L(\bTheta) &=& \sum_{i=1}^n\Bigg[Y_i\Big(\beta_0 + \mathbf{X}_i^\prime \bbeta + \delta \zeta_i \sum_{j=1}^n a_{ij}\mathbf{X}_j^\prime \bbeta\Big)
 - \log\bigg(1+ \exp\Big\{\beta_0 + \mathbf{X}_i^\prime \bbeta + \delta \zeta_i \sum_{j=1}^n a_{ij}\mathbf{X}_j^\prime \bbeta\Big\} \bigg)\nonumber\\
 && +~ \zeta_i (\gamma_0 + \mathbf{X}_i^\prime\bgamma) - \log\big(1+\exp\{\gamma_0 + \mathbf{X}_i^\prime\bgamma\}\big)\Bigg].
\end{eqnarray}
 To estimate the parameter of interest $\bTheta$, we adopt an EM-type algorithm towards the log likelihood function given in (\ref{Eq-3.1}): Let $\mathcal{D}_n= \{(\mathbf{X}_i,Y_i)\}_{i=1}^n$ denote the observed data, and $\hat{\bTheta}^{(k)}$ be the estimate of $\bTheta$ at the $k$th iteration. In the E-step of $(k+1)$th iteration, we calculate the conditional expectation of $L(\bTheta)$ given the observed data $\mathcal{D}_n$ and $\hat{\bTheta}^{(k)}$. i.e.,
\begin{eqnarray}
\ell\big(\bTheta|\hat{\bTheta}^{(k)}\big)&=& E\big\{L(\bTheta)|\mathcal{D}_n, \hat{\bTheta}^{(k)}\big\}\nonumber\\
&=& \underbrace{\sum_{i=1}^nY_i\bigg[\beta_0 + \mathbf{X}_i^\prime \bbeta +  \hat{w}_{ik} \delta\sum_{j=1}^n a_{ij}\mathbf{X}_j^\prime \bbeta\bigg]}_{\ell_1(\bTheta|\hat{\bTheta}^{(k)})} + \ell_2\big(\bTheta|\hat{\bTheta}^{(k)}\big) + \ell_3\big(\bTheta|\hat{\bTheta}^{(k)}\big),
\end{eqnarray}
where $\hat{w}_{ik} = E\{\zeta_i|\mathcal{D}_n, \hat{\bTheta}^{(k)}\}={T_{i1}^{(k)}}/\{T_{i1}^{(k)} + T_{i2}^{(k)}\}$
with
\begin{eqnarray*}
T_{i1}^{(k)} &=& \frac{\exp\bigg\{Y_i\Big(\hat{\beta}^{(k)}_0 + \mathbf{X}_i^\prime \hat{\bbeta}^{(k)} + \hat{\delta}^{(k)}  \sum_{j=1}^n a_{ij}\mathbf{X}_j^\prime \hat{\bbeta}^{(k)}\Big)\bigg\}}{1+ \exp\Big\{\hat{\beta}^{(k)}_0 + \mathbf{X}_i^\prime \hat{\bbeta}^{(k)} + \hat{\delta}^{(k)}  \sum_{j=1}^n a_{ij}\mathbf{X}_j^\prime \hat{\bbeta}^{(k)}\Big\}}\hat{P}_i^{(k)},\\
T_{i2}^{(k)} &=& \frac{\exp\bigg\{Y_i\Big(\hat{\beta}^{(k)}_0 + \mathbf{X}_i^\prime \hat{\bbeta}^{(k)} \Big)\bigg\}}{1+ \exp\Big\{\hat{\beta}^{(k)}_0 + \mathbf{X}_i^\prime \hat{\bbeta}^{(k)}\Big\}}(1-\hat{P}_i^{(k)}),\\
\hat{P}_i^{(k)}&=&   \frac{\exp\Big\{\hat{\gamma}^{(k)}_0 + \mathbf{X}_i^\prime \hat{\bgamma}^{(k)}\Big\}}{1+ \exp\Big\{\hat{\gamma}^{(k)}_0 + \mathbf{X}_i^\prime \hat{\bgamma}^{(k)}\Big\}},
\end{eqnarray*}
and
\begin{eqnarray*}
\ell_2\big(\bTheta|\hat{\bTheta}^{(k)}\big) &=& -\sum_{i=1}^n \hat{w}_{ik} \log\bigg(1+ \exp\Big\{\beta_0 + \mathbf{X}_i^\prime \bbeta + \delta \sum_{j=1}^n a_{ij}\mathbf{X}_j^\prime \bbeta\Big\} \bigg) \\
&&-\sum_{i=1}^n(1-\hat{w}_{ik})\log\big(1+ \exp\{\beta_0 + \mathbf{X}_i^\prime \bbeta\} \big)\\
\ell_3\big(\bTheta|\hat{\bTheta}^{(k)}\big) &=& \sum_{i=1}^n \Big[\hat{w}_{ik} (\gamma_0 + \mathbf{X}_i^\prime\bgamma) - \log\big(1+\exp\{\gamma_0 + \mathbf{X}_i^\prime\bgamma\}\big)\Big].
\end{eqnarray*}

In the M-step of $(k+1)$-th iteration, we can separately maximize the functions $\ell_1\big(\bTheta|\hat{\bTheta}^{(k)}\big) +\ell_2\big(\bTheta|\hat{\bTheta}^{(k)}\big)$ and $\ell_3\big(\bTheta|\hat{\bTheta}^{(k)}\big)$. Because $\ell_3\big(\bTheta|\hat{\bTheta}^{(k)}\big)$ only contains parameters $\gamma_0$ and $\bgamma$, it is straight-forward to find the numerical maximizer $\hat{\bphi}^{(k+1)}=(\hat{\gamma}^{(k+1)}_0,\hat{\bgamma}^{(k+1)\prime})^\prime$ via Newton-Raphson algorithm.
To maximize $\ell_1\big(\cdot|\hat{\bTheta}^{(k)}\big) +\ell_2\big(\cdot|\hat{\bTheta}^{(k)}\big)$  with respect to $\delta$, $\beta_0$ and $\bbeta$, we suggest to use an iterative optimization procedure. Given $\beta_0$ and $\bbeta$, the function $\ell_1\big(\cdot|\hat{\bTheta}^{(k)}\big) +\ell_2\big(\cdot|\hat{\bTheta}^{(k)}\big)$ a univariate concave function of $\delta$, which can be easily maximized by Newton-Raphson method. Denote
\begin{eqnarray*}
\hat{\delta}^{(k+1)}= \arg\max_{\delta} \Big\{\ell_1\big(\delta,\hat{\beta}_0^{(k)},\hat{\bbeta}^{(k)}|\hat{\bTheta}^{(k)}\big) +\ell_2\big(\delta,\hat{\beta}_0^{(k)},\hat{\bbeta}^{(k)}|\hat{\bTheta}^{(k)}\big)\Big\}.
\end{eqnarray*}
In order to guarantee algorithmic stability, we specify $\delta = \hat{\delta}^{(k+1)}$ and $\bbeta=\hat{\bbeta}^{(k)}$ in the term $\delta\sum_{j=1}^n a_{ij}\mathbf{X}_j^\prime \bbeta$ of $\ell_1\big(\cdot|\hat{\bTheta}^{(k)}\big) +\ell_2\big(\cdot|\hat{\bTheta}^{(k)}\big)$, which leads to profiled log likelihood function of $\beta_0$ and $\bbeta$:
\begin{eqnarray*}
L_{pro}(\beta_0,\bbeta)&=&\sum_{i=1}^nY_i(\beta_0 + \mathbf{X}_i^\prime \bbeta +  \hat{w}_{ik}\hat{u}_{ik})
 -\sum_{i=1}^n \hat{w}_{ik} \log\bigg(1+ \exp\Big\{\beta_0 + \mathbf{X}_i^\prime \bbeta + \hat{u}_{ik}\Big\} \bigg) \\
&&-\sum_{i=1}^n(1-\hat{w}_{ik})\log\big(1+ \exp\{\beta_0 + \mathbf{X}_i^\prime \bbeta\} \big),
\end{eqnarray*}
where $\hat{u}_{ik}=\hat{\delta}^{(k+1)}\sum_{j=1}^n a_{ij}\mathbf{X}_j^\prime \hat{\bbeta}^{(k)}$. Denote $(\hat{\beta}_0^{(k+1)}, \hat{\bbeta}^{(k+1)})$ as the  maximizer of $L_{pro}(\beta_0,\bbeta)$. In the proposed EM algorithm, we set the initial estimators of the parameters as follows: $\hat{\delta}^{(0)} =0$, $\hat{\gamma}_0^{(0)} = 0$, $ \hat{\bgamma}^{(0)} = \mathbf{0}$, $\hat{\beta}_0^{(0)}$ and $ \hat{\bbeta}^{(0)}$ are the maximum likelihood estimators of classical logistic regression. We repeatedly apply the E-step and M-step until convergence with $\|\hat{\bTheta}^{(k+1)}-\hat{\bTheta}^{(k)}\| < 10^{-6}$. The final estimator is denoted as $\hat{\bTheta} = (\hat{\delta}, \hat{\beta}_0, \hat{\bbeta}^\prime, \hat{\gamma}_0,\hat{\bgamma}^\prime)^\prime$. The performance of EM-based estimator will be carefully studied via numerical simulation. Here we point out that the term $\hat{w}_{ik}$ at convergence
indicates the posterior probability that the $i$th node might be affected by its neighbor's
behavior. For $i=1,\cdots,n$, we denote $\hat{w}_{i} = {T_{i1}^{}}/\{T_{i1}^{} + T_{i2}^{}\}$, where
\begin{eqnarray*}
T_{i1}^{} &=& \frac{\exp\bigg\{Y_i\Big(\hat{\beta}^{}_0 + \mathbf{X}_i^\prime \hat{\bbeta}^{} + \hat{\delta}^{}  \sum_{j=1}^n a_{ij}\mathbf{X}_j^\prime \hat{\bbeta}^{}\Big)\bigg\}}{1+ \exp\Big\{\hat{\beta}^{}_0 + \mathbf{X}_i^\prime \hat{\bbeta}^{} + \hat{\delta}^{}  \sum_{j=1}^n a_{ij}\mathbf{X}_j^\prime \hat{\bbeta}^{}\Big\}}\hat{P}_i^{},\\
T_{i2}^{} &=& \frac{\exp\bigg\{Y_i\Big(\hat{\beta}^{}_0 + \mathbf{X}_i^\prime \hat{\bbeta}^{} \Big)\bigg\}}{1+ \exp\Big\{\hat{\beta}^{}_0 + \mathbf{X}_i^\prime \hat{\bbeta}^{}\Big\}}(1-\hat{P}_i^{}),\\
\hat{P}_i^{}&=&   \frac{\exp\Big\{\hat{\gamma}^{}_0 + \mathbf{X}_i^\prime \hat{\bgamma}^{}\Big\}}{1+ \exp\Big\{\hat{\gamma}^{}_0 + \mathbf{X}_i^\prime \hat{\bgamma}^{}\Big\}}.
\end{eqnarray*}
We call $\hat{w}_{i}$ as the posterior ``susceptible" probability for the $i$th node. A larger $\hat{w}_{i}$
indicates that the outcome of $i$th node is more likely to be  by its neighbor's
feature. In this case, the predicted outcome $\hat{Y}_i$ satisfies:
\begin{align}
\mathbb{{P}}(\hat{Y}_i = 1|\mathbf{X}_i, \hat{\zeta}_i) &= \frac{\exp\{\hat{\beta}_0 + \mathbf{X}_i^\prime \hat{\bbeta} + \hat{\delta} \hat{\zeta}_i \sum_{j=1}^n a_{ij}\mathbf{X}_j^\prime \hat{\bbeta}\}}{ 1+ \exp\{\hat{\beta}_0 + \mathbf{X}_i^\prime \hat{\bbeta} + \hat{\delta} \hat{\zeta}_i \sum_{j=1}^n a_{ij}\mathbf{X}_j^\prime \hat{\bbeta}\}},~~ i=1,\cdots,n,
\end{align}
where $\hat{\delta}$, $\hat{\beta}_0$ and $\hat{\bbeta}$ are the EM-based estimator; $\hat{\zeta}_i$ is generated from the Binomial distribution with $\mathbb{P}(\hat{\zeta}_i = 1) = \hat{w}_{i}$, $i=1,\cdots,n$.

\begin{table}[H]
 \begin{center}
{\bf Table 1.} Type I error and power of the proposed test with significance level $\alpha=0.05$$^\dagger$.
{\rule{128mm}{0.13mm}\\
\begin{tabular}{llccccccccc}
 &&$\delta = 0$ &&$\delta = 0.01$&& $\delta = 0.03$ && $\delta = 0.05$&& $\delta = 0.10$\\
 \hline
 Case I  && 0.048 &&  0.116  && 0.802 && 0.996&& 1\\
 Case II && 0.048 &&  0.098  && 0.794 && 0.992&& 1\\
\end{tabular}\\
\rule{128mm}{0.13mm}}\\
\end{center}
\end{table}

\section{Simulation Study}
\setcounter{equation}{0}
In this section, we conduct some simulations to verify the performance of our proposed method.   We generate the graph (network) from the  stochastic block model \cite[]{SBM-1983}. Specifically, we assume there are five communities in the network, and the stochastic block model is given by
\begin{eqnarray*}
\mathbb{P}(a_{ij} =1) = 1-\mathbb{P}(a_{ij} =0) = \mathbf{P}_{C_iC_j},
\end{eqnarray*}
where $\mathbf{P}$ is a $5\times 5$ symmetric matrix whose $(C_i,C_j)$th element $\mathbf{P}_{C_iC_j}$
denotes the probability that communities $C_i$ and $C_j$ are connected.  The total number of nodes is set as $n=2000$, and the numbers of nodes contained in each community are $(500, 500, 400, 400, 200)$. The probability matrix $\mathbf{P}$ has elements $\mathbf{P}_{11} = 0.01$, $\mathbf{P}_{22} = \mathbf{P}_{55}=0.10$, $\mathbf{P}_{33}=0.05$, $\mathbf{P}_{44} = 0.15$, and $\mathbf{P}_{C_iC_j} = 10^{-4}$ for $C_i\neq C_j$.
The covariates $\mathbf{X}_i = (X_{i1}, X_{i2})^\prime$ are generated from two situations:

{\it Case I}: $X_{i1}$ follows from a Bernoulli distribution with the success probability 0.5, and $X_{i2}$
follows from a uniform distribution over $(-1,1)$.

{\it Case II}: $X_{i1}$ follows from $N(0,1)$ and $X_{i2}$ follows from $N(0,2)$.\\
We set the parameters $\beta_0 = 0.5$, $\beta_1 = -1$, $\beta_2 = 1$, $\gamma_0 = 0$, $\gamma_1 = -1$ and $\gamma_2 = 1$. The computation is implemented with the help of R software. All the simulation results are based on 500 replicates.

\begin{table}[H]
\begin{center}
 {{\bf Table 2}. The Bias and MSE of estimators with proposed method$^\dagger$.}
\vspace{0.3cm}

\small
\begin{tabular}{cccccccccccccc}
\hline
  &&   \multicolumn{2}{c}{$\delta = 0.1$} &  & \multicolumn{2}{c}{$\delta = 0.3$} \\
\cline{3-4}\cline{6-7}
 && Case I & Case II  &&  Case I & Case II \\
\hline
$\delta$ && (0.0035, 0.0013)   & (0.0022, 0.0003)   &&  (0.0031, 0.0098)  &(0.0011, 0.0005)   \\
$\gamma_0$ && (0.0870, 0.1652)   & (0.0188, 0.2502)   &&  (-0.0302, 0.0527)  &(-0.0014, 0.0327)   \\
$\gamma_1$ && (0.0791, 0.3190)   & (0.1645, 0.5752)   &&  (0.0075, 0.0985)  &(0.0094,  0.0427)   \\
$\gamma_2$ && (-0.0604, 0.3102)   & (-0.1209, 0.3559)   &&  (0.0193, 0.1046)  &(-0.0169, 0.0245)   \\
$\beta_0$ && (0.0034, 0.0178)   & (0.0043, 0.0046)   &&  (-0.0180, 0.0425)  &(-0.0067, 0.0058)   \\
$\beta_1$ && (-0.0059, 0.0289)   & (-0.0049, 0.0058)   &&  (-0.0224, 0.0392)  &(-0.0069,  0.0078)   \\
$\beta_2$ && (-0.0017, 0.0235)   & (0.0043, 0.0030)   &&  (0.0335, 0.0505)  &(0.0083,  0.0042)   \\
\hline
\end{tabular}
\end{center}
\end{table}

The first simulation  is to validate the effectiveness of the perturbed test statistic given in (\ref{PTS-3.5}).
We set the  values of $\delta$ to be 0, 0.01, 0.03, 0.05 and 0.10, respectively.   The p-value of the test statistic is calculated by generating 1000 perturbed test statistics as described in Section 3.  In Table 1, we report the empirical type I error and power of proposed testing method with significance level $\alpha = 0.05$.
From the results in Table 1, we can see that the type I error is reasonable under the null when  $\delta = 0$.
Moreover, the power of the test is increasing as $\delta$ increases. Therefore, the proposed testing procedure works well in terms of diagnosing the existence  of network in logistic regression model.

\begin{table}[ht]
\begin{center}
 {{\bf Table 3}. The Bias and MSE of estimators with  the classical logistic regression model$^\dagger$.}
\vspace{0.3cm}

\small
\begin{tabular}{cccccccccccccc}
\hline
  &&   \multicolumn{2}{c}{$\delta = 0.1$} &  & \multicolumn{2}{c}{$\delta = 0.3$} \\
\cline{3-4}\cline{6-7}
 && Case I & Case II  &&  Case I & Case II \\
\hline
$\beta_0$ && (-0.6432, 0.4182)   & (0.0158, 0.0092)   &&  (-0.9797, 0.9647)  &(0.0529, 0.0164)   \\
$\beta_1$ && (-0.1353, 0.0302)   & (0.1022, 0.0162)   &&  (-0.2028, 0.0534)  &(0.3232, 0.1092)   \\
$\beta_2$ && (0.1309, 0.0255)   & (-0.1028, 0.0135)    &&  (0.2063, 0.0525)  &(-0.3224, 0.1067)   \\
\hline
\end{tabular}
\end{center}
\end{table}
We perform the second simulation to investigate the performance of EM estimation algorithm under $\delta \neq 0$. The generation of data is the same with the first simulation, except that $\delta$ is chosen as 0.1 and 0.3, respectively. In Table 2, we report the simulation results of proposed estimator, including the bias (Bias) given by the difference of sample means of estimator and the true value, and the mean squared error (MSE) of estimator.
The results in Table 2 indicate that all estimators are unbiased, except for $\gamma_1$ and $\gamma_2$ with $\delta = 0.1$ (Case II). One possible explanation for this phenomenon is due to the fact that $\bphi=(\gamma_0,\bgamma^\prime)^\prime$ is unidentified for small $\delta$. In addition, the performance of $\gamma$ with $\delta = 0.3$ is much better than that of $\delta = 0.1$. For comparison, we also consider the following  classical logistic regression model when modeling the binary outcome:
\begin{align}\label{Logi-5.3}
\mathbf{P}(Y_i = 1|\mathbf{X}_i) &= \frac{\exp\{\beta_0 + \mathbf{X}_i^\prime \bbeta\}}{ 1+ \exp\{\beta_0 + \mathbf{X}_i^\prime \bbeta\}},~~ i=1,\cdots,n.
\end{align}
Note that the  classical logistic model (\ref{Logi-5.3}) does not describe the influence of network among nodes (samples). In Table 3, we report the Bias and MSE  for maximum likelihood estimators of $\beta_0$ and $\bbeta$ with model (\ref{Logi-5.3}), where the data are generated as the first simulation with $\delta$ = 0.1 and 0.3. The results in Table 3 indicate that the estimators are biased if we ignore the network of samples. Basically, the proposed method works well when there exists potential network among samples with binary outcomes.

\begin{figure}[H]
  \centering
  \begin{subfigure}{0.45\textwidth}
    \includegraphics[width=\textwidth]{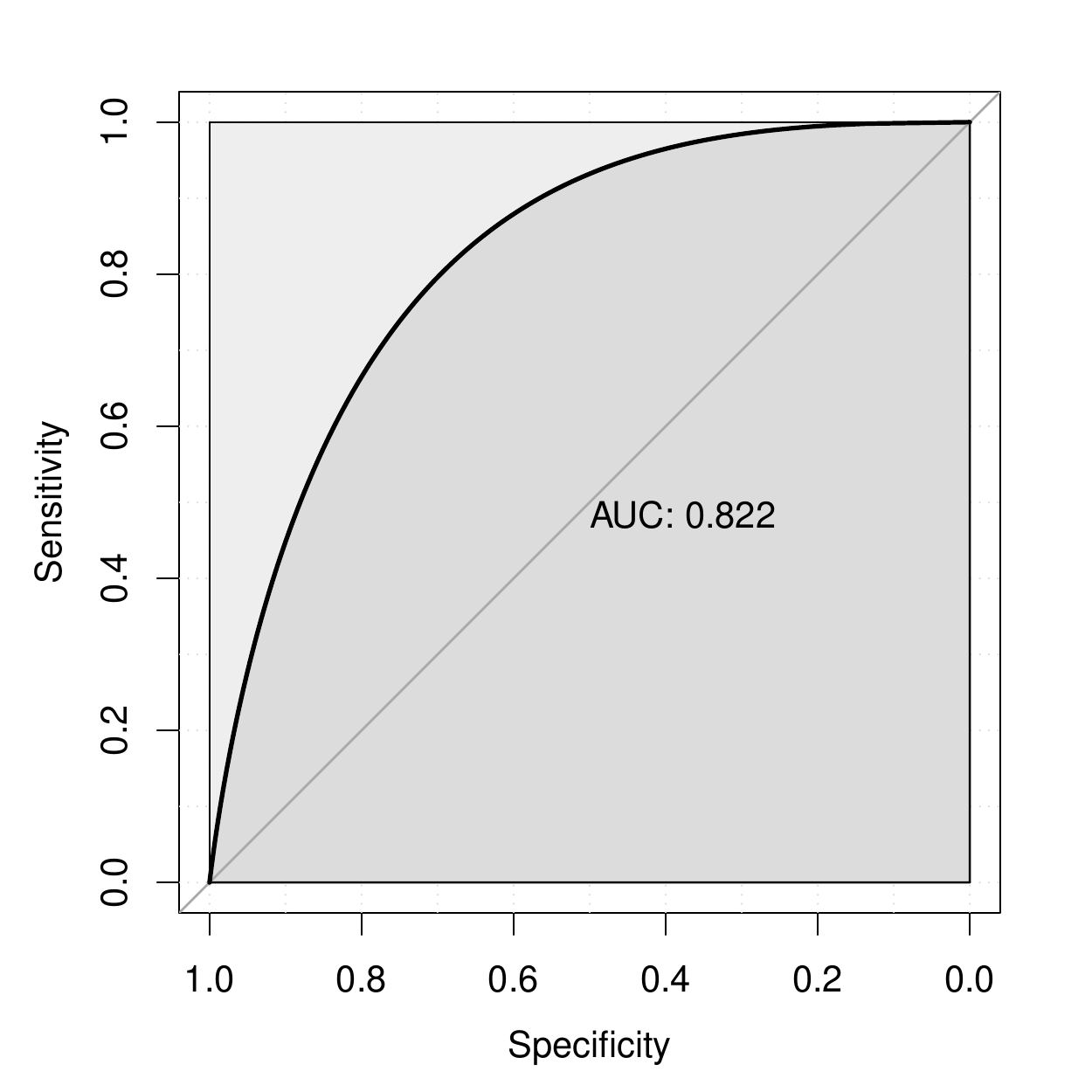}
    \caption{ROC of proposed method}
  \end{subfigure}
  \begin{subfigure}{0.45\textwidth}
    \includegraphics[width=\textwidth]{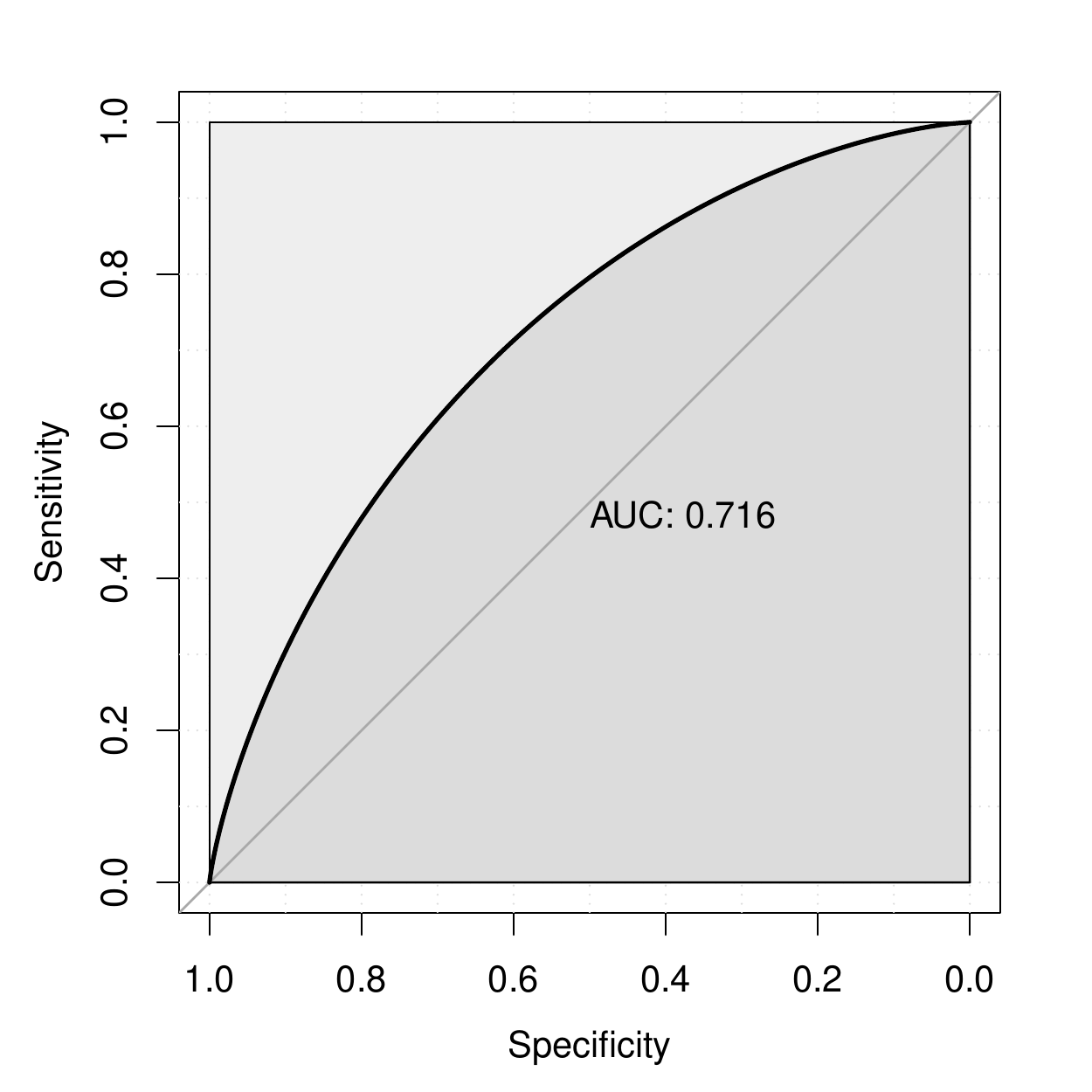}
   \caption{ROC of logistic model}
  \end{subfigure}
 \vspace{-0.1cm}
\begin{center}
{{\bf Figure 1}. {The ROC plots with Case I and $\delta = 0.1$ in the simulation}.}
\end{center}
\end{figure}
\begin{figure}[H]
  \centering
  \begin{subfigure}{0.45\textwidth}
    \includegraphics[width=\textwidth]{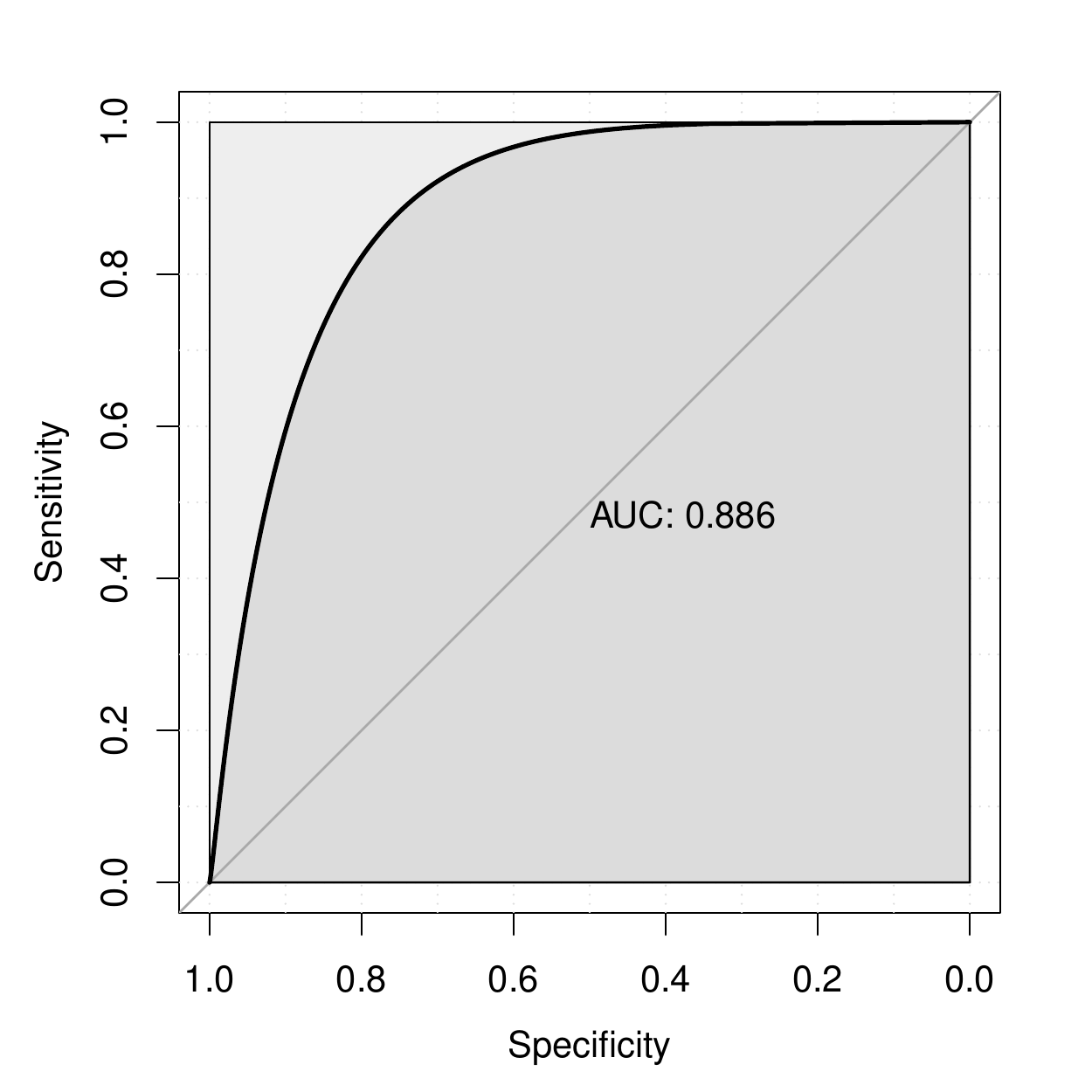}
    \caption{ROC of proposed method}
  \end{subfigure}
  \begin{subfigure}{0.45\textwidth}
    \includegraphics[width=\textwidth]{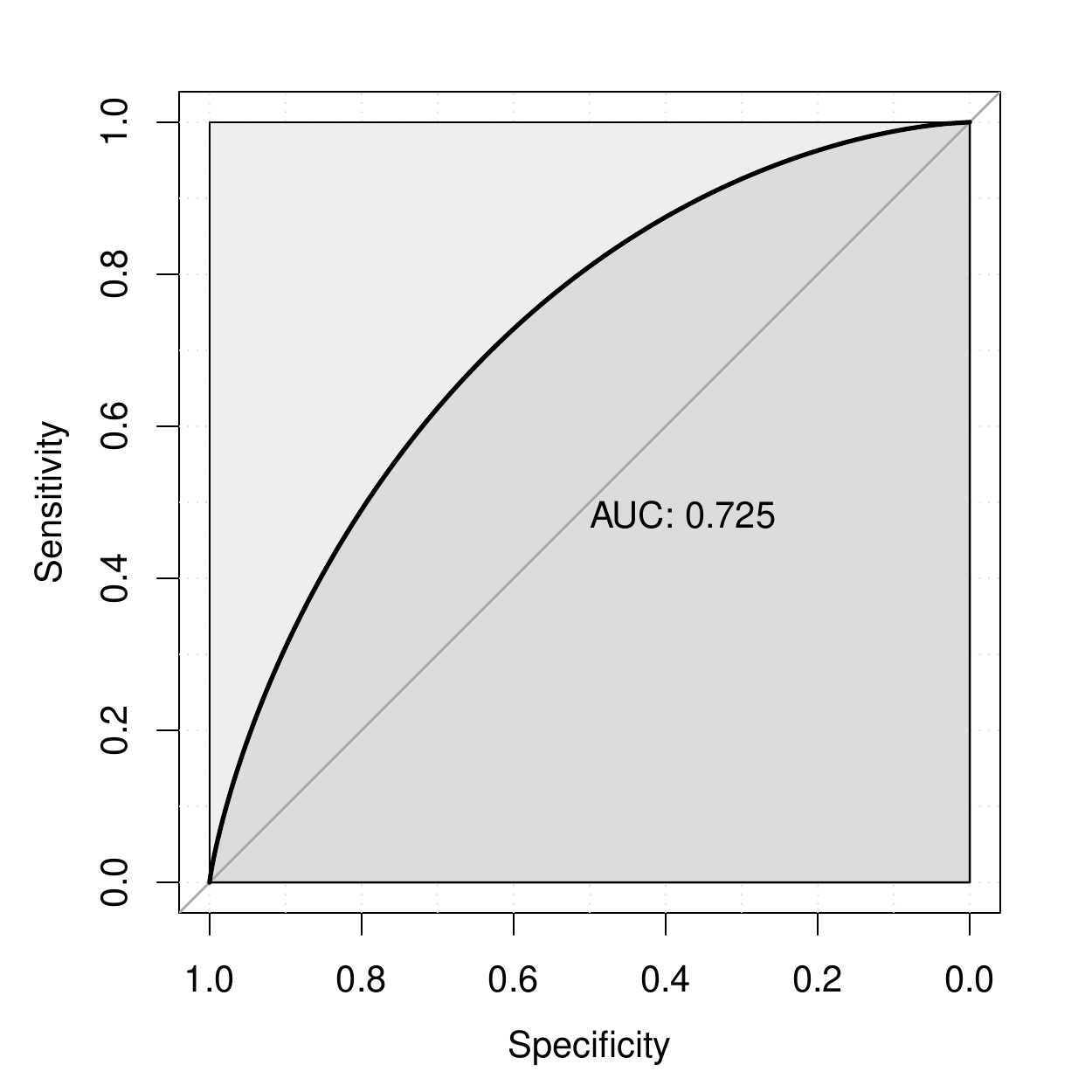}
   \caption{ROC of logistic model}
  \end{subfigure}
 \vspace{-0.1cm}
\begin{center}
{{\bf Figure 2}. {The ROC plots with Case I and $\delta = 0.3$ in the simulation}.}
\end{center}
\end{figure}

\begin{figure}[H]
  \centering
  \begin{subfigure}{0.45\textwidth}
    \includegraphics[width=\textwidth]{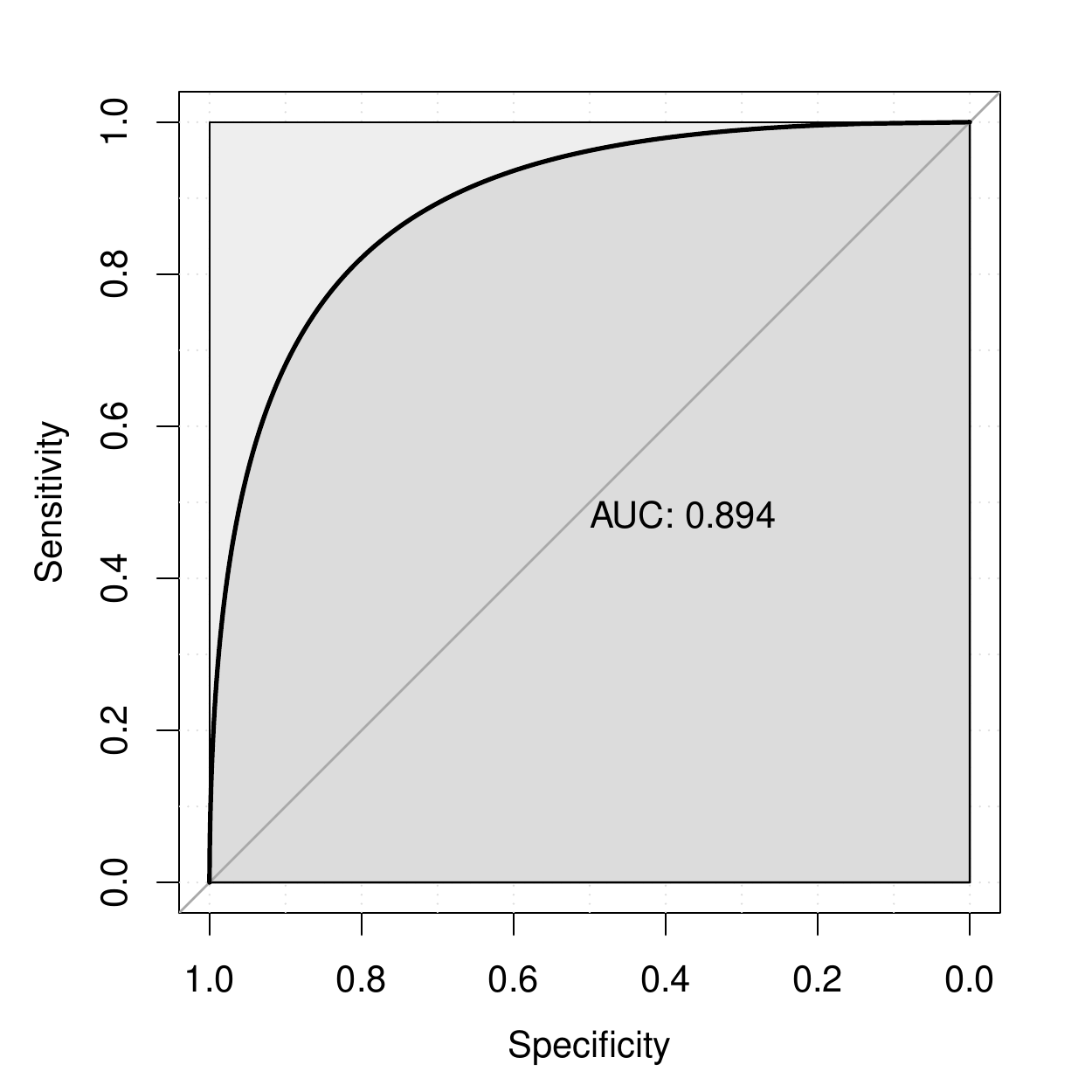}
    \caption{ROC of proposed method}
  \end{subfigure}
  \begin{subfigure}{0.45\textwidth}
    \includegraphics[width=\textwidth]{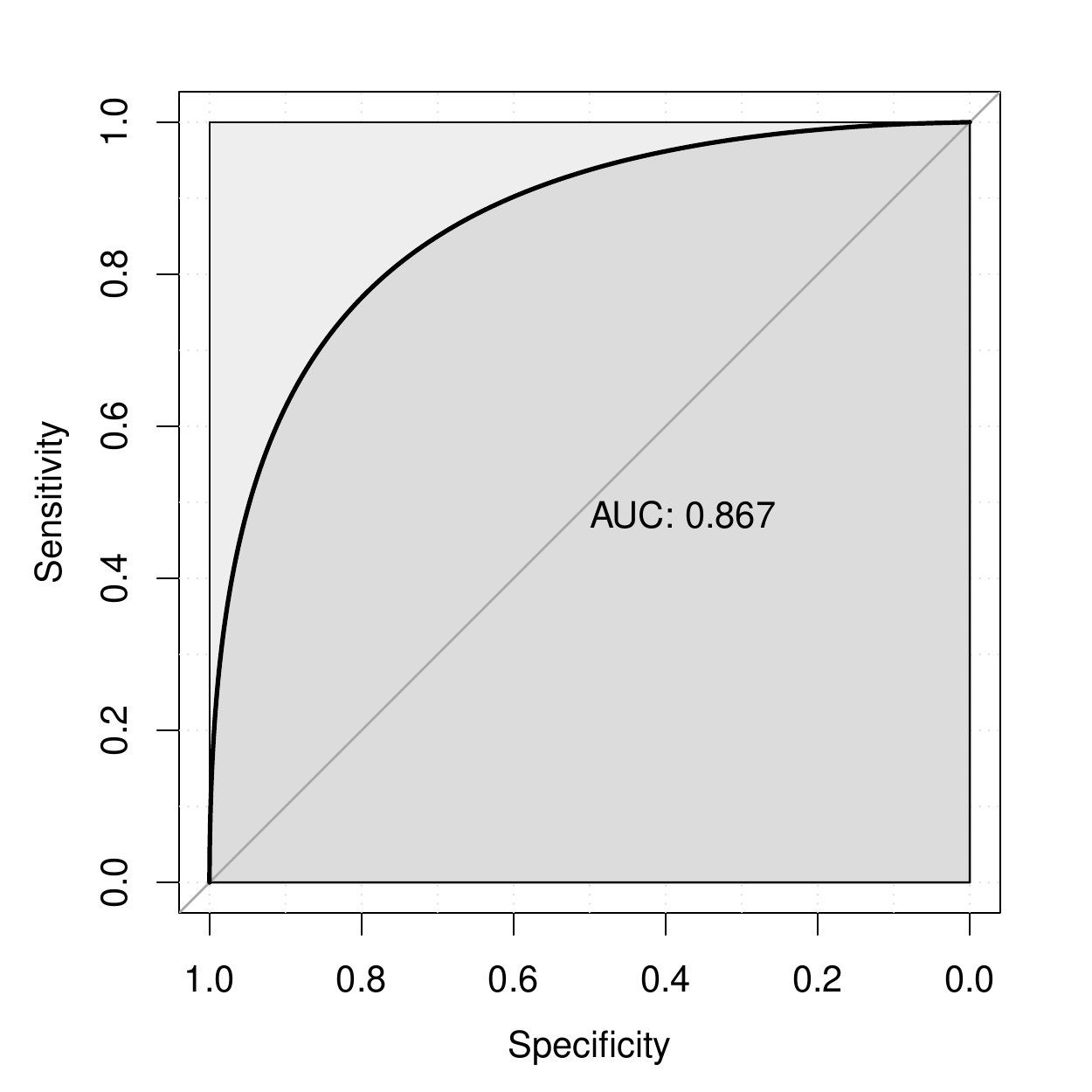}
   \caption{ROC of logistic model}
  \end{subfigure}
 \vspace{-0.1cm}
\begin{center}
{{\bf Figure 3}. {The ROC plots with Case II and $\delta = 0.1$ in the simulation}.}
\end{center}
\end{figure}

\begin{figure}[H]
  \centering
  \begin{subfigure}{0.45\textwidth}
    \includegraphics[width=\textwidth]{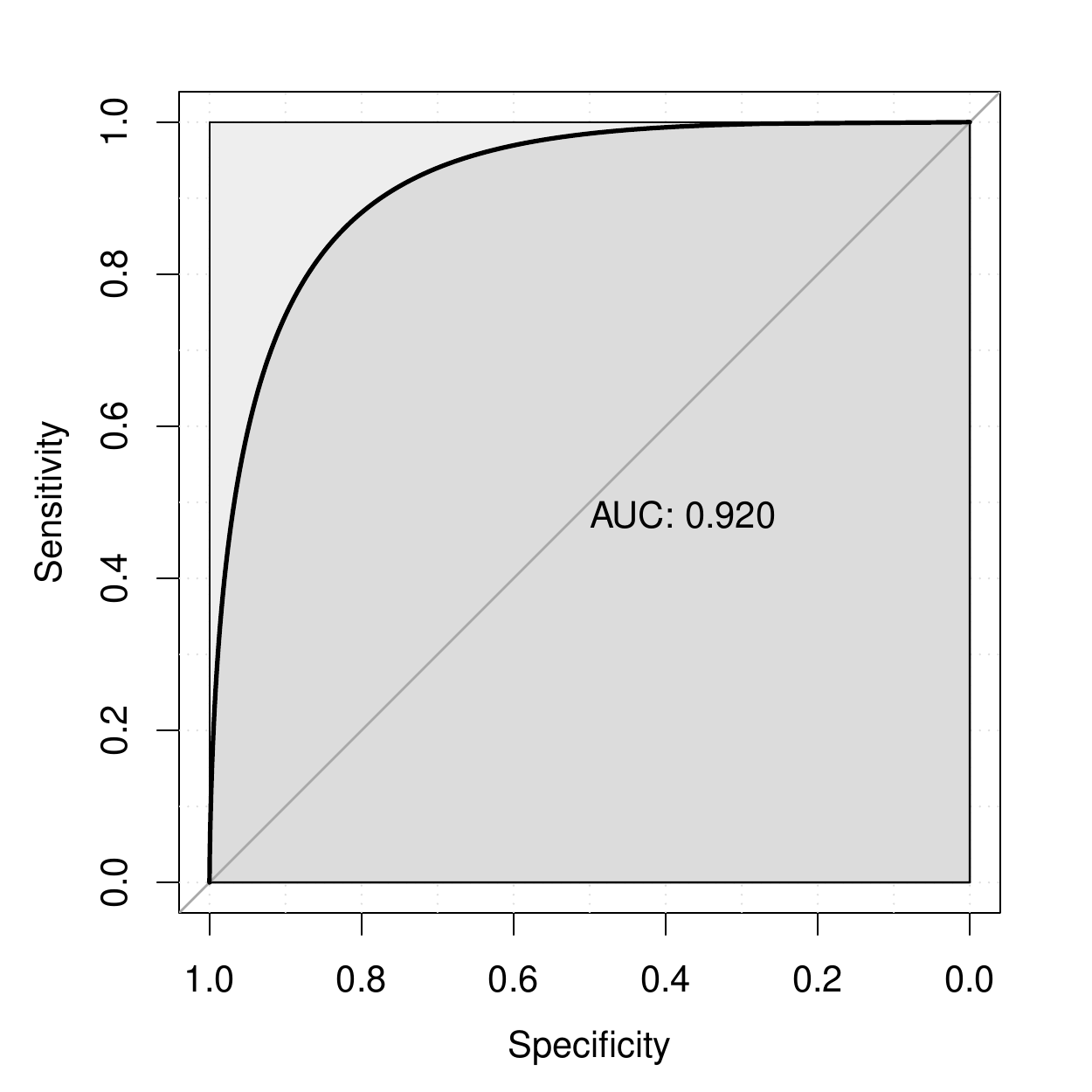}
    \caption{ROC of proposed method}
  \end{subfigure}
  \begin{subfigure}{0.45\textwidth}
    \includegraphics[width=\textwidth]{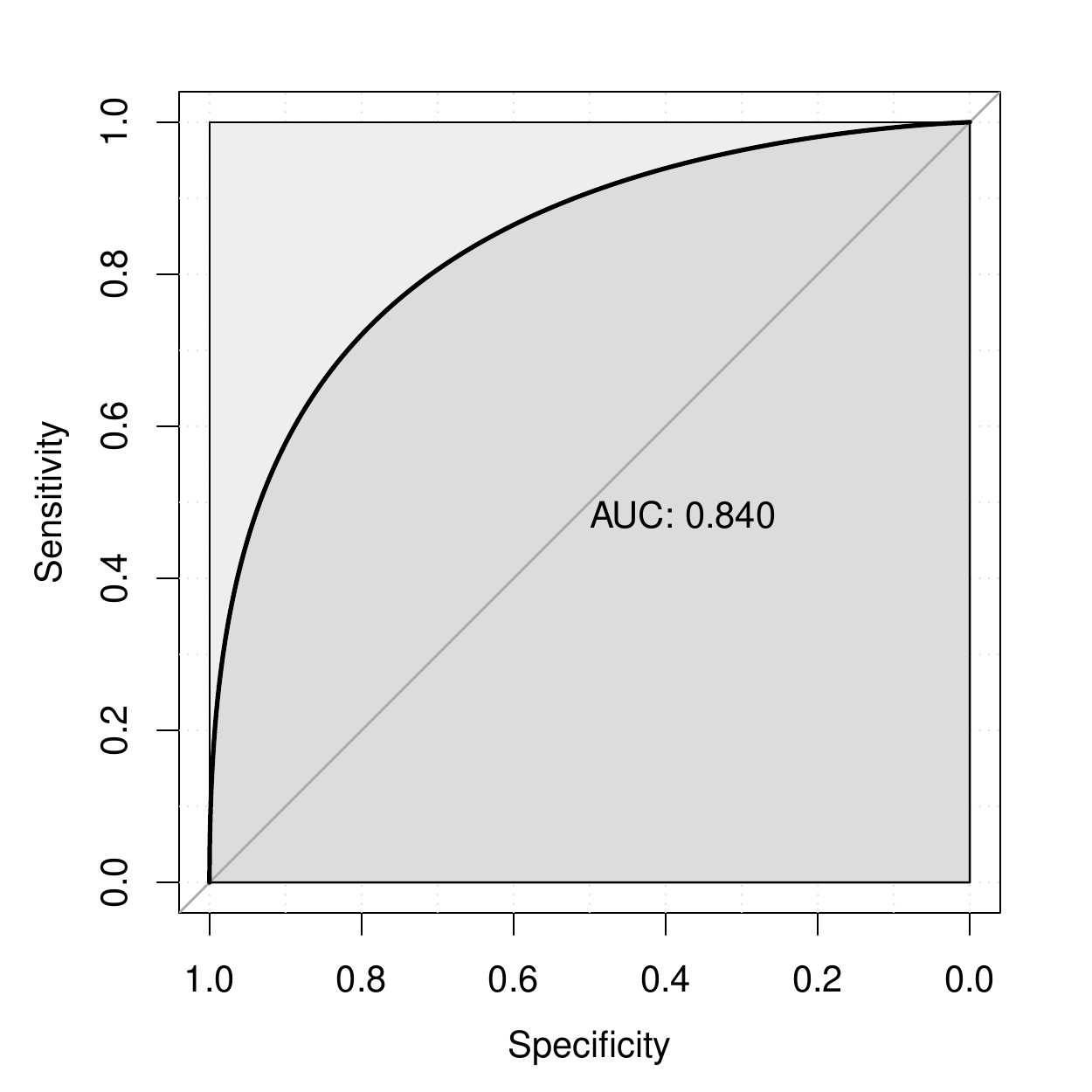}
   \caption{ROC of logistic model}
  \end{subfigure}
 \vspace{-0.1cm}
\begin{center}
{{\bf Figure 4}. {The ROC plots with Case II and $\delta = 0.3$ in the simulation}.}
\end{center}
\end{figure}

The receiver operating characteristic (ROC) curve is a very useful tool  evaluate classifiers in many practical applications \cite[]{rocA-2006}. In Figures 1-2, we report the ROC curves with $\delta = 0.1$ (other cases are similar and omitted), which are plotted by the R package {\tt pROC} \cite[]{pROC-2011}. From the view of ROC analysis, the area under the curve (AUC) measures the performance of a classifier. A larger value of  AUC
means a better classification capacity. In Figures 1-2 we also present the AUCs of two  classifiers, which indicate the proposed method has a larger AUC compared with the classical logistic regression model. i.e., it is useful to use the network information when predicting the binary outcomes with our proposed model.

\begin{figure}[H]
\begin{center}
\includegraphics[width=3in]{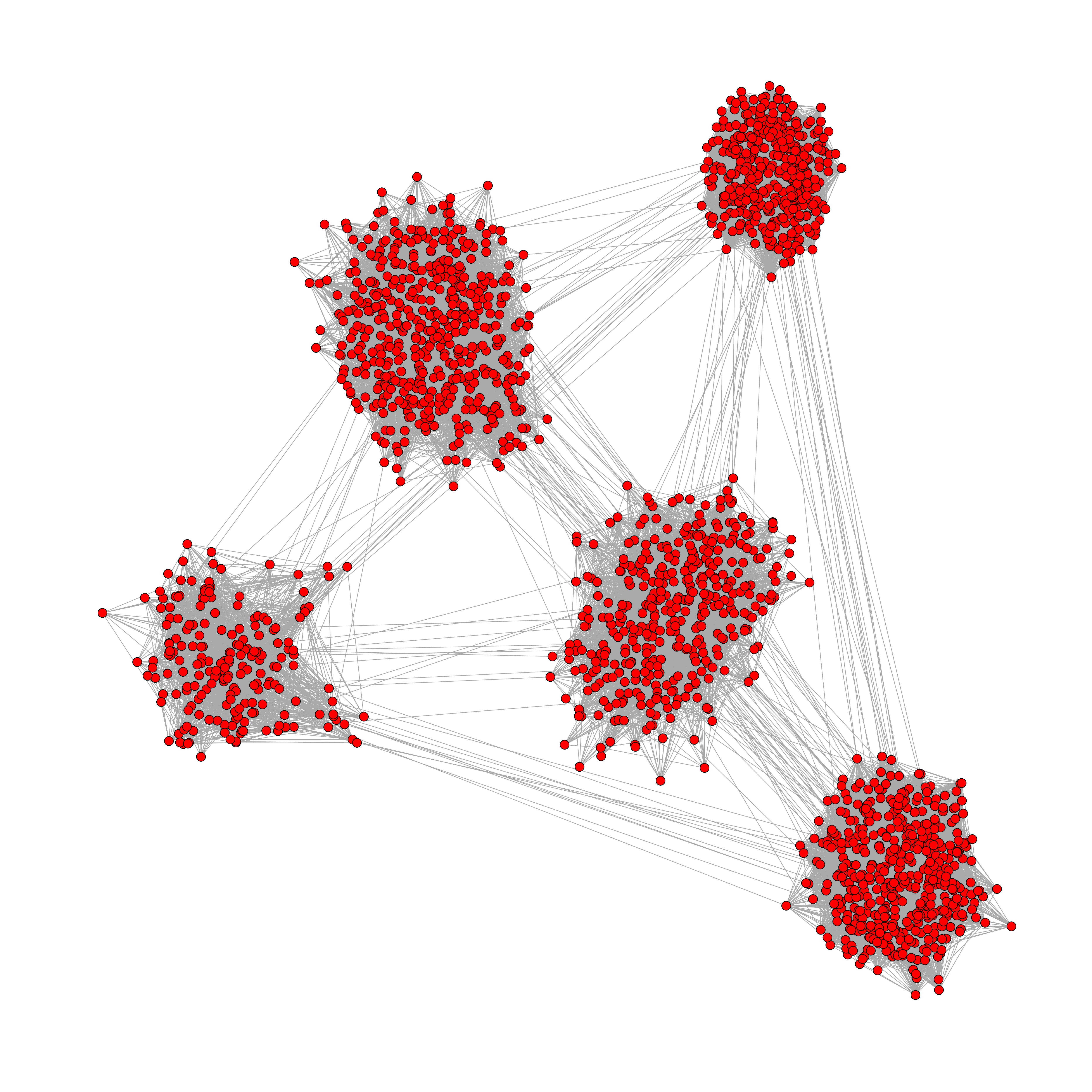}\\
\vspace{-0.5cm}
\footnotesize{\bf Figure 5.} { The plot of network for survival data in the Section 6.}
\end{center}
\end{figure}



\section{Application}
\subsection{A generated network with survival data}
In this section, we apply our proposed method to a generated network with survival data, which is  available in the online supplemental material of \cite{su-JASA-2020}. There are total 2000 nodes in the network ($n=2000$), and each node contains two covariates ($X_{i1}$ and $X_{i2}$), the observed failure time and censored indicator, where $X_{i1}$ is binary (e.g. male =1, female=0), $X_{i2}$ is continuous (e.g. age), $i=1,\cdots,n$. Denote the response $Y_i=1$ if the $i$th node is not censored, and $Y_i=0$ otherwise. The second variable $X_2$ is scaled with zero mean and unit variance. We use the proposed models (\ref{M-1}) and (\ref{M-2}) to analyse this network dataset. In Figure 3, we illustrate the above-mentioned network data in this application. It is clear that there are five  communities in this network.

\begin{table}[H]
\begin{center}
 {{\bf Table 4}. The summary of parameter estimators in the Section 6.1$^\dagger$.}
\vspace{0.3cm}
\small
\begin{tabular}{lccccccccccccc}
\hline
 &$\delta$ &&$\gamma_0$ &&$\gamma_1$ &&$\gamma_2$&&$\beta_0$ &&$\beta_1$ &&$\beta_2$ \\
\hline
Proposed &0.0275 && -0.2861 &&  3.3878 && -1.7049 &&0.0495 &&1.1521 &&-0.6966 \\
Logistic &$*$ && $*$ &&$*$ && $*$ &&0.3143 &&1.4415 &&-0.8353 \\
\hline
\end{tabular}
\end{center}
{\vspace{-0.2cm}  \hspace{0.2cm}\tiny $\dagger$ ``Proposed" denotes our proposed method; ``Logistic" denotes the the classical logistic regression model; ``$*$" denotes the value is not available.}
\end{table}

\begin{figure}[H]
  \centering
  \begin{subfigure}{0.45\textwidth}
    \includegraphics[width=\textwidth]{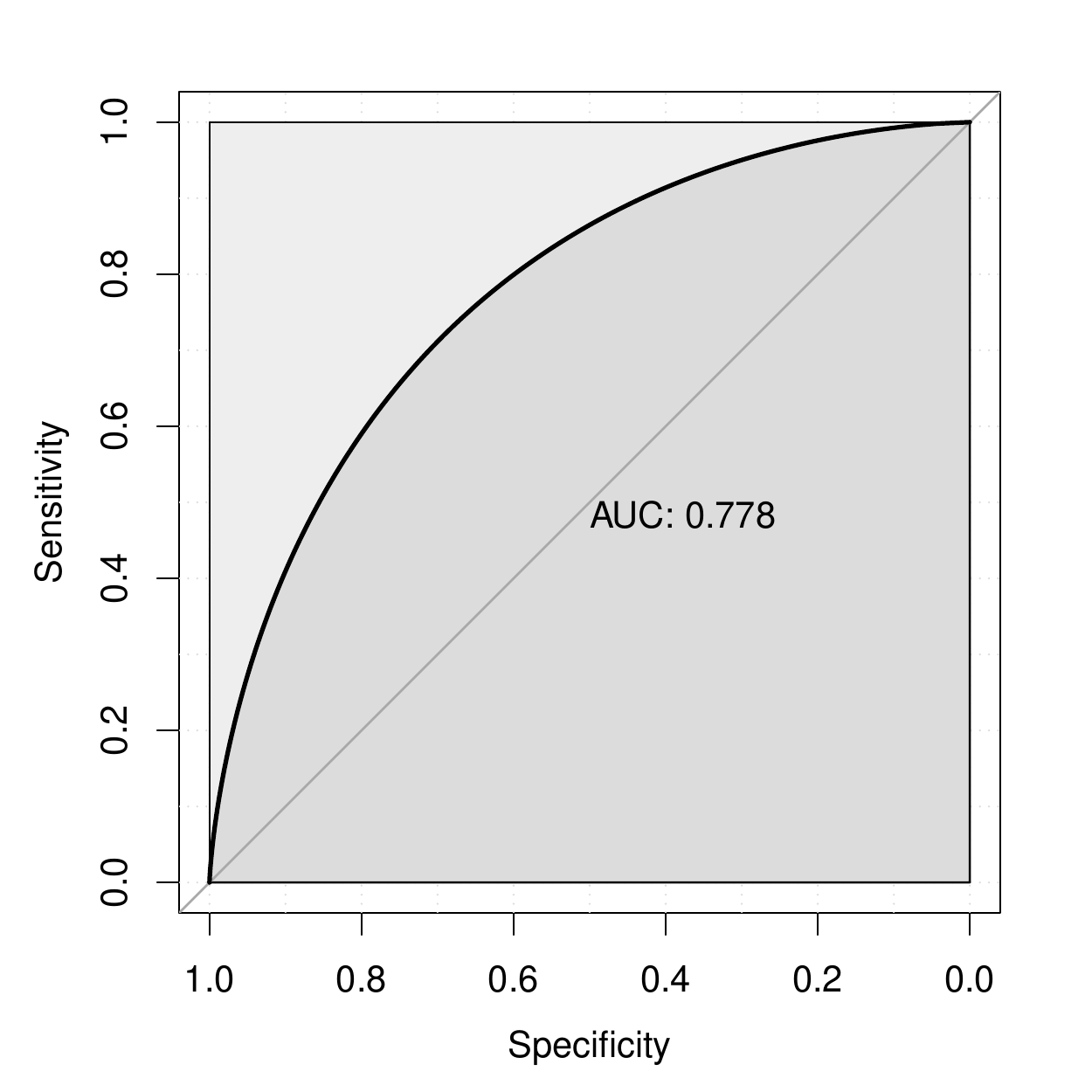}
    \caption{ROC of proposed method}
  \end{subfigure}
  \begin{subfigure}{0.45\textwidth}
    \includegraphics[width=\textwidth]{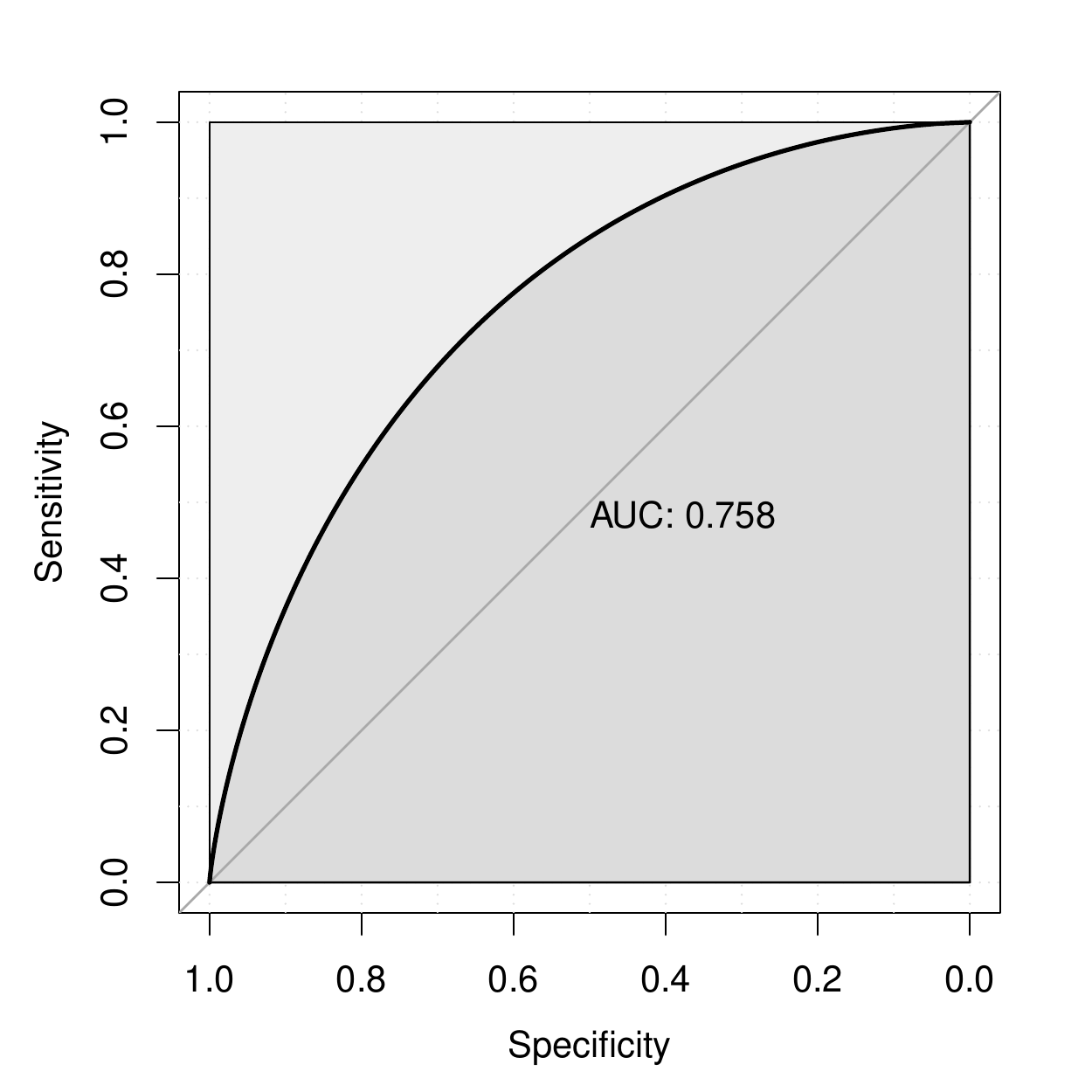}
   \caption{ROC of logistic model}
  \end{subfigure}
 \vspace{-0.1cm}
\begin{center}
{{\bf Figure 6}. {The ROC plots in the application (Section 6.1)}.}
\end{center}
\end{figure}
First we employ the proposed score test for  $H_0: \delta =0$, where the perturbed test statistic $T_n^*$ is calculated 1000 times. At significance level $\alpha = 0.05$,  the empirical upper $\alpha$-quantile, $C_\alpha =9.44$, and the value of test statistic $T_n$ is 19.02.  In view of the fact that $T_n > C_\alpha $, we can reject the null hypothesis $H_0$.  i.e., there exists network dependence among nodes' binary outcomes. Next, we estimate the model parameters by the proposed EM  algorithm, where the results are presented in Table 4. The estimated parameter $\delta$ is 0.0275,  indicating a significant positive  network dependence among those
susceptible nodes. For comparison, we also use the classical logistic model to fit this dataset, where the estimated parameters are given in Table 5. To evaluate the performance of classification, we report the ROC curves of both methods in Figure 4. Because the AUC of proposed method is larger than that of classical logistic model, our method is desirable to fit this network dataset. In addition, we calculate the estimated posterior ``susceptible" probability for each node, where the  histogram of the estimated posterior is reported in Figure 5.
The majority of ``susceptible" probabilities are large, indicating that those nodes are more likely to be influenced by their neighbors in the network.

\begin{figure}[H]
\begin{center}
\includegraphics[width=4in]{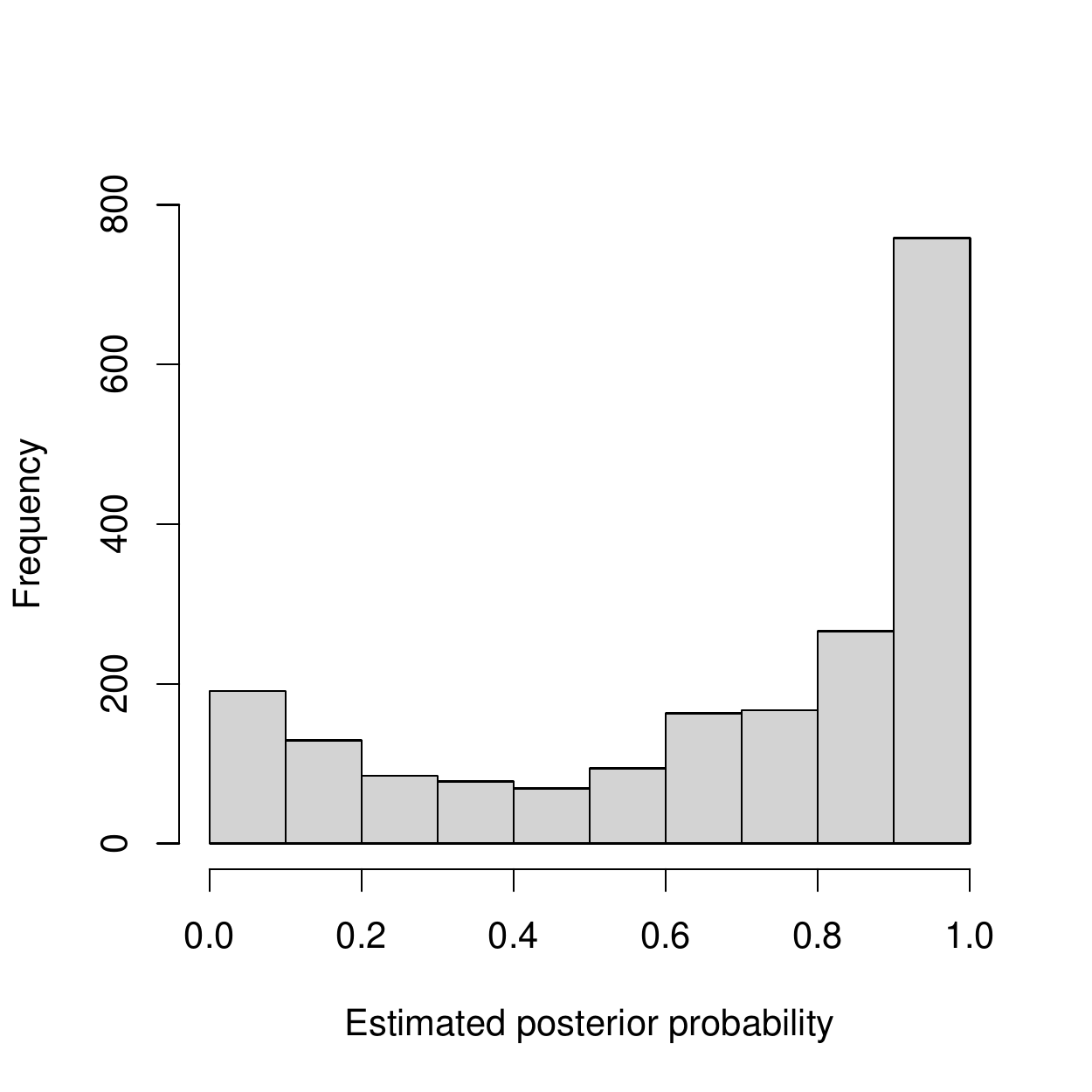}\\
\vspace{-0.5cm}
\footnotesize{\bf Figure 7.} { Histogram of the estimated susceptible probabilities for the nodes in Section 6.1.}
\end{center}
\end{figure}


\begin{figure}[H]
\begin{center}
\includegraphics[width=4in]{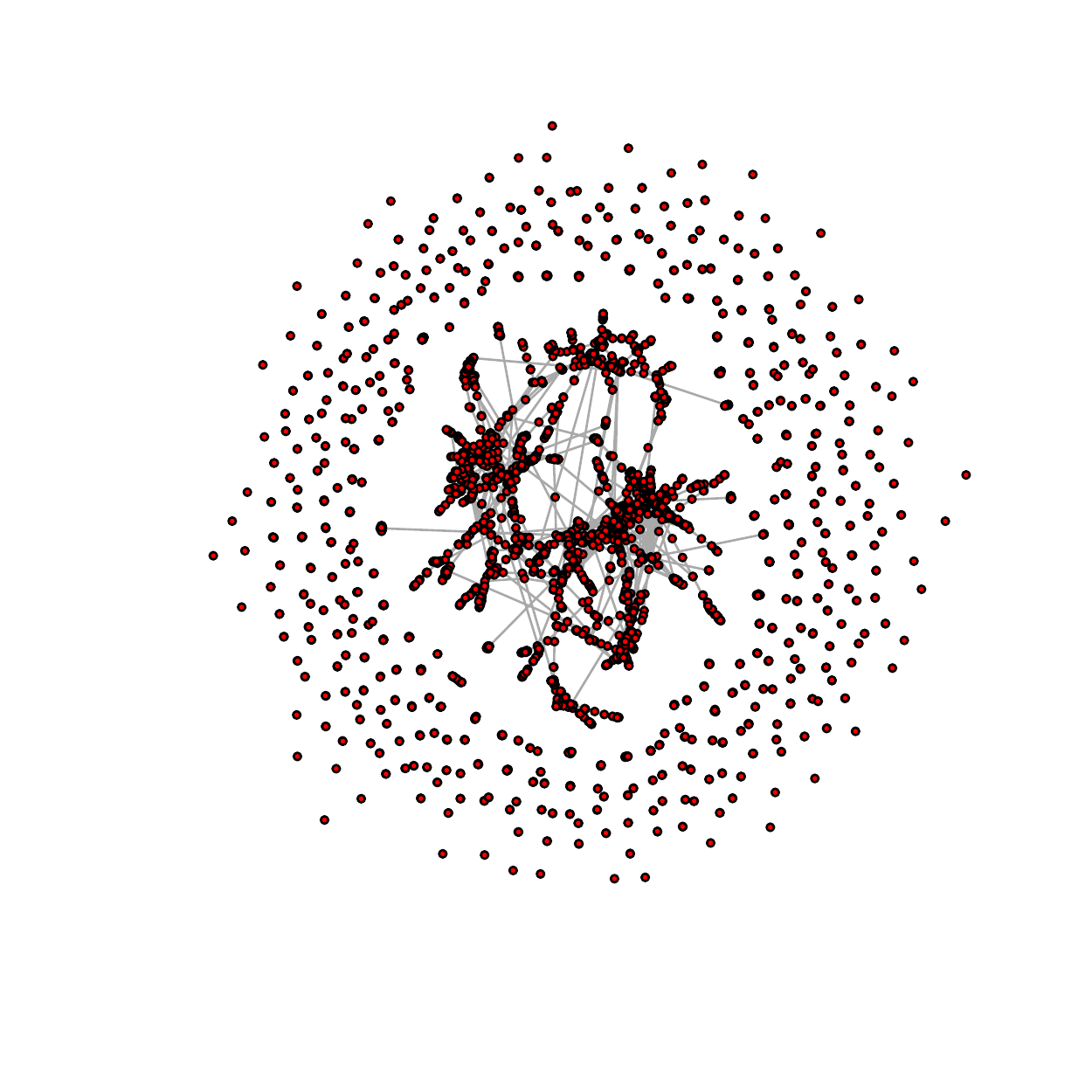}\\
\vspace{-1cm}
\footnotesize{\bf Figure 8.} { The plot of network for CiteSeer dataset in the Section 6.2.}
\end{center}
\end{figure}

\subsection{The CiteSeer dataset with citation network}
In this section, we apply our proposed method to the CiteSeer dataset,  consisting of 3312 scientific publications classified into one of six classes.  The datset is public available at https://linqs.soe.ucsc.edu/data.
Every node (publication) in the dataset is described by a 0/1-valued word vector (features) indicating the absence/presence of the corresponding word from the dictionary with 3703 unique words. We use the PCA dimension reduction to obtain two covariates ($X_{i1}$ and $X_{i2}$) for each node.  Denote the response $Y_i=1$ if the $i$th node is about artificial intelligence, and $Y_i=0$ otherwise. We use the proposed models (\ref{M-1}) and (\ref{M-2}) to analyse this network dataset. In Figure 8, we plot the CiteSeer dataset network, indicating this network is very sparse.

\begin{table}[H]
\begin{center}
 {{\bf Table 5}. The summary of parameter estimators in the Section 6.2$^\dagger$.}
\vspace{0.3cm}
\small
\begin{tabular}{lccccccccccccc}
\hline
 &$\delta$ &&$\gamma_0$ &&$\gamma_1$ &&$\gamma_2$&&$\beta_0$ &&$\beta_1$ &&$\beta_2$ \\
\hline
Proposed &1.8986 &&-1.0855 && -3.9585 && 6.1909 &&-2.4419 &&0.3034 &&-0.1540 \\
Logistic &$*$ && $*$ &&$*$ && $*$ &&-2.5526 &&0.3985 &&-0.3590\\
\hline
\end{tabular}
\end{center}
{\vspace{-0.2cm}  \hspace{0.2cm}\tiny $\dagger$ ``Proposed" denotes our proposed method; ``Logistic" denotes the the classical logistic regression model; ``$*$" denotes the value is not available.}
\end{table}

\begin{figure}[H]
  \centering
  \begin{subfigure}{0.45\textwidth}
    \includegraphics[width=\textwidth]{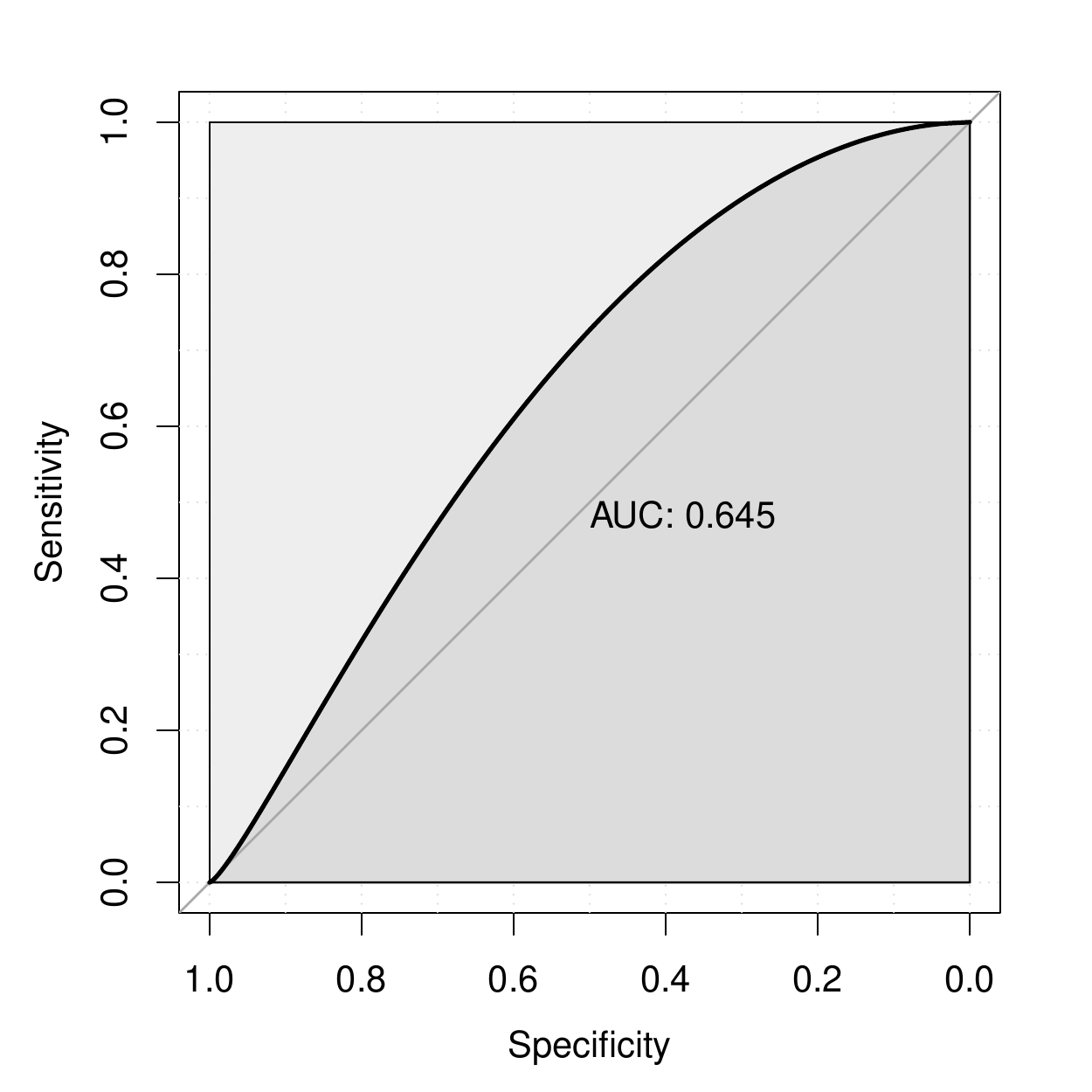}
    \caption{ROC of proposed method}
  \end{subfigure}
  \begin{subfigure}{0.45\textwidth}
    \includegraphics[width=\textwidth]{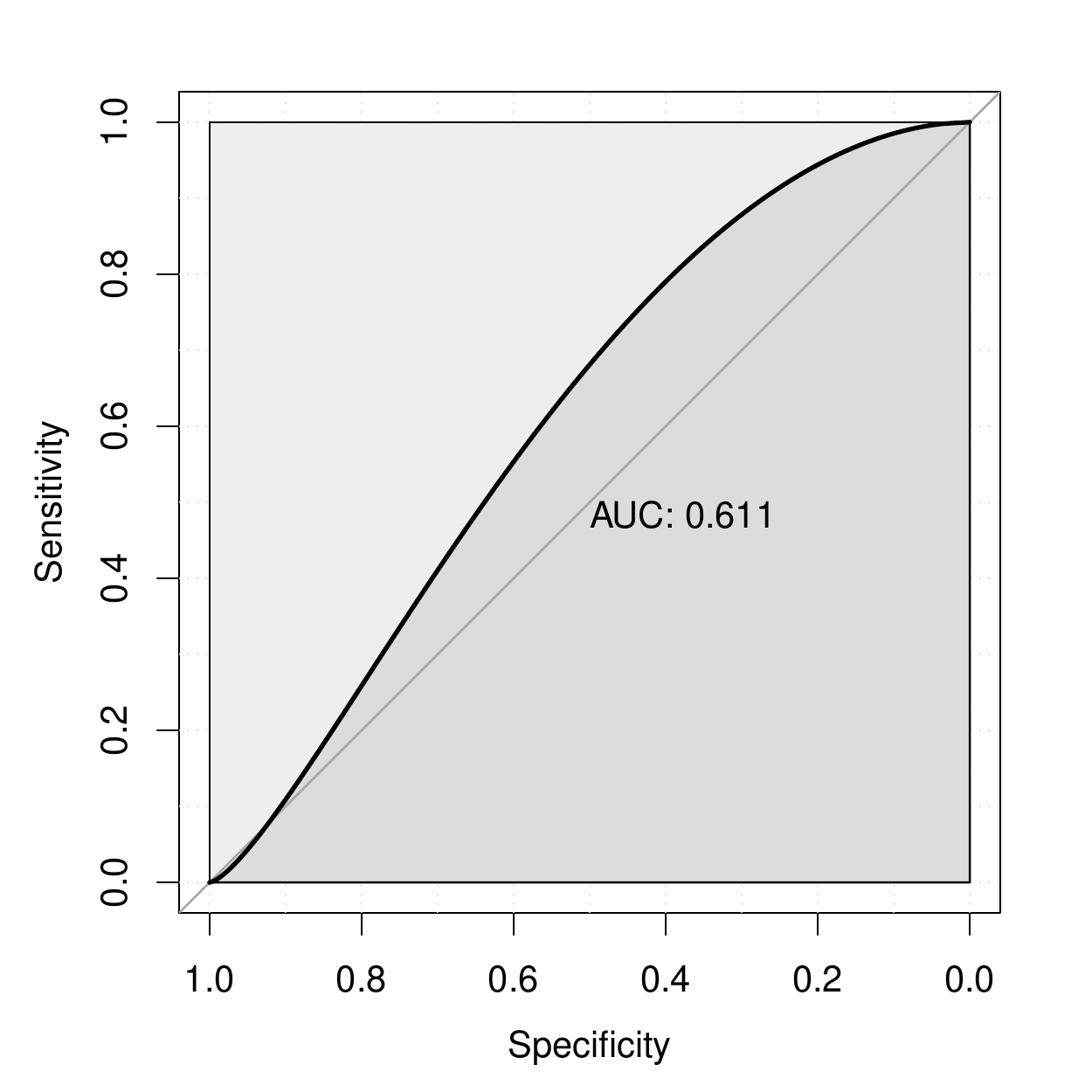}
   \caption{ROC of logistic model}
  \end{subfigure}
 \vspace{-0.1cm}
\begin{center}
{{\bf Figure 9}. {The ROC plots in the application (Section 6.2)}.}
\end{center}
\end{figure}
First we use the proposed score test for  $H_0: \delta =0$, where the perturbed test statistic $T_n^*$ is calculated 1000 times. With significance level $\alpha = 0.05$,  the empirical upper $\alpha$-quantile, $C_\alpha = 9.01$, and the value of test statistic $T_n$ is 41.20.  Due to the fact that $T_n > C_\alpha $, we can reject the null hypothesis $H_0$.  i.e., there exists network dependence among nodes' binary outcomes. Next, we estimate the model parameters by the proposed EM  algorithm, where the results are presented in Table 5. The estimated parameter $\delta$ is 1.8986,  indicating a significant positive  network dependence among those
susceptible nodes. For comparison, we also use the classical logistic model to fit this dataset, where the estimated parameters are given in Table 6. To evaluate the performance of classification, we report the ROC curves of both methods in Figure 9. Because the AUC of proposed method is larger than that of classical logistic model, our method is desirable to fit this network dataset. Furthermore, we calculate the estimated posterior ``susceptible" probability for each node, where the  histogram of the estimated posterior is reported in Figure 10.
The majority of ``susceptible" probabilities are low for most nodes, but can be high for a group of node.

\begin{figure}[H]
\begin{center}
\includegraphics[width=4in]{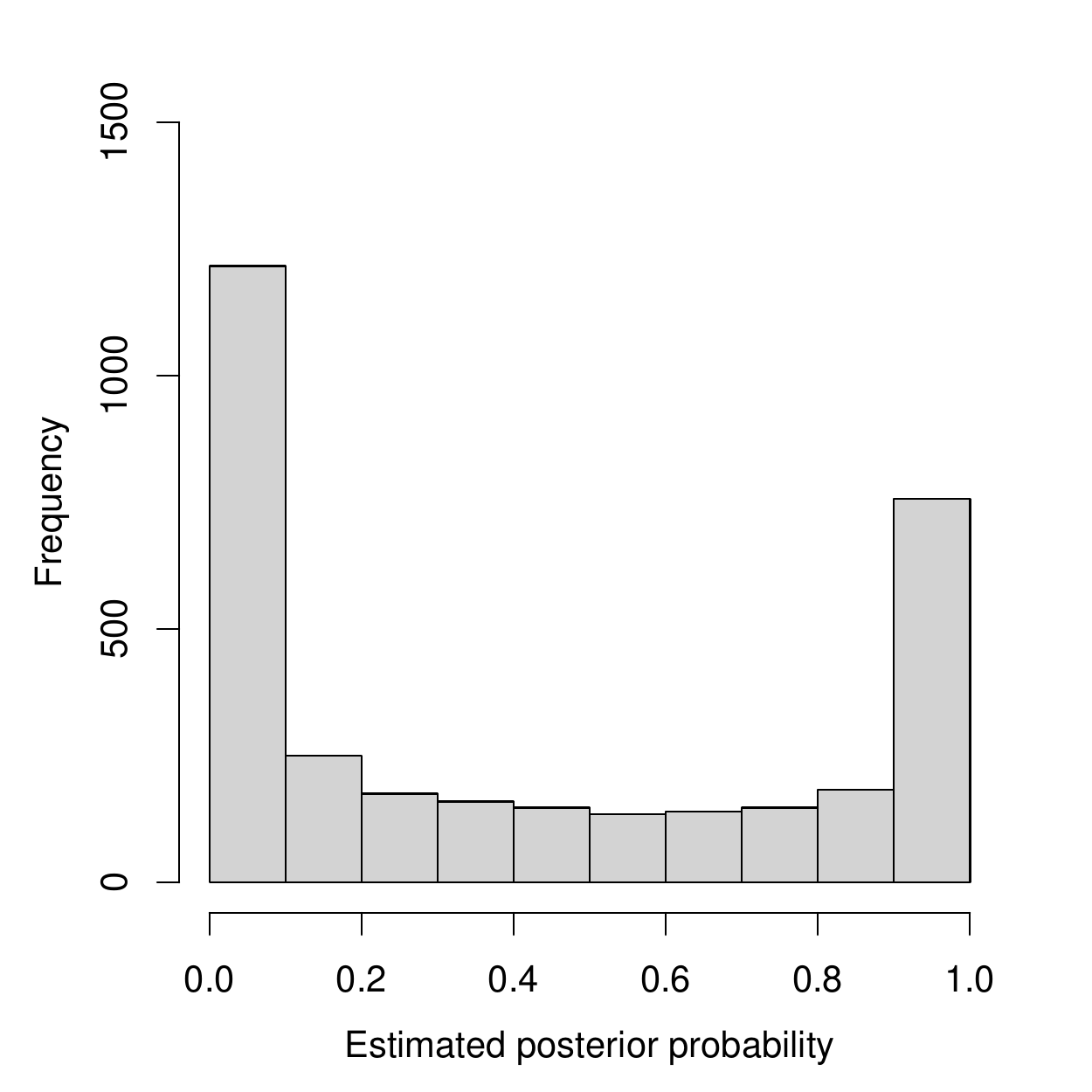}\\
\vspace{-0.1cm}
\footnotesize{\bf Figure 10.} { Histogram of the estimated susceptible probabilities for the nodes in Section 6.2.}
\end{center}
\end{figure}

\section{Concluding Remarks}

In this paper, we have proposed a novel latent  logistic regression model to describe network dependence. We introduced a  score-type test to check the existence of network dependence.  An EM-type algorithm was used to estimate the model parameters. Simulations and applications were used to validate the usefulness of the proposed method.  It is interesting to extend our idea to multi-class logistic regression model with graph data, which will be studied in the future of our research.

\vspace{1cm}
\section*{Appendix}
\setcounter{equation}{0}
{\bf Proof of Theorem 1:} Note that $\tilde{\bfeta} =(\tilde{\beta}_0,\tilde{\bbeta}^\prime)^\prime$ is the standard maximum likelihood estimator with $\delta =0$, where the asymptotic property of $\tilde{\bfeta}$ has been well established. By the Taylor expansion, we have
\begin{eqnarray*}
\frac{1}{\sqrt{n}}S^*(\tilde{\btheta};\bphi) = \frac{1}{\sqrt{n}}S^*(\btheta_t;\bphi) - \mathcal{B}^*_n(\btheta_t)
\mathcal{I}_n^{-1}(\btheta_t)\frac{1}{\sqrt{n}}S_1(\btheta_t;\bphi) + o_P(1),
\end{eqnarray*}
where
\begin{eqnarray*}
S_1(\btheta_t;\bphi) &=& \frac{\partial L(\btheta;\bzeta)}{\partial \bfeta}|_{\btheta = \btheta_t}\\
 &= & \sum_{i=1}^n \left[Y_i - \frac{\exp(\beta_{0t}+\mathbf{X}_i^\prime\bbeta_t)}{1+\exp(\beta_{0t}+\mathbf{X}_i^\prime\bbeta_t)}\right](1,\mathbf{X}_i^\prime)^\prime.
\end{eqnarray*}
Hence, some calculations lead to the following expression:
\begin{eqnarray*}
\frac{1}{\sqrt{n}}S^*(\tilde{\btheta};\bphi) &=& \frac{1}{\sqrt{n}}\sum_{i=1}^n \tilde{Z}^*_i\left[Y_i - \frac{\exp(\beta_{0t}+\mathbf{X}_i^\prime\bbeta_t)}{1+\exp(\beta_{0t}+\mathbf{X}_i^\prime\bbeta_t)}\right]\\
 &&- \mathcal{B}^*_n(\btheta_t)\mathcal{I}_n^{-1}(\btheta_t)\frac{1}{\sqrt{n}}\sum_{i=1}^n \left[Y_i - \frac{\exp(\beta_{0t}+\mathbf{X}_i^\prime\bbeta_t)}{1+\exp(\beta_{0t}+\mathbf{X}_i^\prime\bbeta_t)}\right](1,\mathbf{X}_i^\prime)^\prime + o_P(1)\\
 &=& \frac{1}{\sqrt{n}}\sum_{i=1}^n U_i^*(\btheta_t;\bphi) + o_P(1).
\end{eqnarray*}

Under some regularity conditions, it can be proved that $\sigma^2(\bphi) = \lim_{n\rightarrow \infty}n^{-1}\sum_{i=1}^n \{U_i^*(\btheta_t;\bphi)\}^2$ is finite and can be consistently estimated by $n^{-1}\sum_{i=1}^n \{U_i^*(\tilde{\btheta};\bphi)\}^2$. For fixed $\bphi$, we get that as $n\rightarrow \infty$,
\begin{eqnarray*}
n^{-1/2}S^*(\tilde{\btheta};\bphi)\stackrel{d}{\longrightarrow} N(0,\sigma^2(\bphi)),
\end{eqnarray*}
where $\stackrel{d}{\longrightarrow}$ denotes convergence in distribution. Moreover, we can derive that $n^{-1/2}S^*(\tilde{\btheta};\bphi)$ converges in distribution to a Gaussian process with
mean 0 and covariance matrix \\$\lim_{n\rightarrow \infty} n^{-1}\sum_{i=1}^n U_i^*(\btheta_t;\bphi_1)U_i^*(\btheta_t;\bphi_2)$ for any $\bphi_1$, $\bphi_2\in \bGamma$. This ends the proof.

\bibliographystyle{natbib}
\bibliography{reference}

\begin{thebibliography}{}

\bibitem[Carroll and Pederson(1993)]{Carroll-JRSSB-1993}
Carroll, R.~J. and Pederson, S. (1993).
\newblock On robustness in the logistic regression model.
\newblock \emph{Journal of the Royal Statistical Society: Series B
  (Methodological)} \textbf{55}, 693--706.

\bibitem[Chandna \emph{et~al.}(2021)Chandna, Olhede, and Wolfe]{Chandna-2021}
Chandna, S., Olhede, S., and Wolfe, P. (2021).
\newblock Local linear graphon estimation using covariates.
\newblock \emph{Biometrika}  DOI:10.1093/biomet/asab057.

\bibitem[Conroy and Sajda(2012)]{Bryan-PMLR-2012}
Conroy, B. and Sajda, P. (2012).
\newblock Fast, exact model selection and permutation testing for
  l2-regularized logistic regression.
\newblock \emph{Proceedings of the Fifteenth International Conference on
  Artificial Intelligence and Statistics} \textbf{22}, 246--254.

\bibitem[Das \emph{et~al.}(2013)Das, Moore, Wong, Stumpf, Oberst, McIntosh, and
  Burnett]{AI-S_Das-2013}
Das, S., Moore, T., Wong, W., Stumpf, S., Oberst, I., McIntosh, K., and
  Burnett, M. (2013).
\newblock End-user feature labeling: Supervised and semi-supervised approaches
  based on locally-weighted logistic regression.
\newblock \emph{Artificial Intelligence} \textbf{204}, 56--74.

\bibitem[Efron(1988)]{Efron-JASA-1988}
Efron, B. (1988).
\newblock Logistic regression, survival analysis, and the kaplan-meier curve.
\newblock \emph{Journal of the American statistical Association} \textbf{83},
  414--425.

\bibitem[Fan \emph{et~al.}(2017)Fan, Song, and Lu]{FanSL-JASA-2017}
Fan, A., Song, R., and Lu, W. (2017).
\newblock Change-plane analysis for subgroup detection and sample size
  calculation.
\newblock \emph{Journal of the American Statistical Association} \textbf{112},
  769--778.

\bibitem[Fawcett(2006)]{rocA-2006}
Fawcett, T. (2006).
\newblock An introduction to roc analysis.
\newblock \emph{Pattern Recognition Letters} \textbf{27}, 861--874.

\bibitem[Han \emph{et~al.}(2019)Han, Hong, Cheon, and Park]{Kyoohyung-22019}
Han, K., Hong, S., Cheon, J.~H., and Park, D. (2019).
\newblock Logistic regression on homomorphic encrypted data at scale.
\newblock \emph{Proceedings of the AAAI Conference on Artificial Intelligence}
  \textbf{33}, 9466--9471.

\bibitem[Holland \emph{et~al.}(1983)Holland, Laskey, and Leinhardt]{SBM-1983}
Holland, P., Laskey, K., and Leinhardt, S. (1983).
\newblock Stochastic blockmodels: First steps.
\newblock \emph{Social networks} \textbf{5}, 109--137.

\bibitem[Jennings(1986)]{Jennings-1986-JASA}
Jennings, D. (1986).
\newblock Outliers and residual distributions in logistic regression.
\newblock \emph{Journal of the American Statistical Association} \textbf{81},
  987--990.

\bibitem[Landwehr \emph{et~al.}(1984)Landwehr, Pregibon, and
  Shoemaker]{Landwehr-JASA-1984}
Landwehr, J.~M., Pregibon, D., and Shoemaker, A.~C. (1984).
\newblock Graphical methods for assessing logistic regression models.
\newblock \emph{Journal of the American Statistical Association} \textbf{79},
  61--71.

\bibitem[Lemon \emph{et~al.}(2003)Lemon, Roy, Clark, Friedmann, and
  Rakowski]{Stephenie-2003}
Lemon, S., Roy, J., Clark, M., Friedmann, P., and Rakowski, W. (2003).
\newblock Classification and regression tree analysis in public health:
  Methodological review and comparison with logistic regression.
\newblock \emph{Annals of Behavioral Medicine} \textbf{26}, 172--181.

\bibitem[Li \emph{et~al.}(2019)Li, Bellotti, and Adams]{Yazhe-2019}
Li, Y., Bellotti, T., and Adams, N. (2019).
\newblock Issues using logistic regression with class imbalance, with a case
  study from credit risk modelling.
\newblock \emph{Foundations of Data Science} \textbf{1}, 389--417.

\bibitem[Liu \emph{et~al.}(2009)Liu, Chen, and Ye]{jun-2009}
Liu, J., Chen, J., and Ye, J. (2009).
\newblock Large-scale sparse logistic regression.
\newblock \emph{Proceedings of the 15th ACM SIGKDD international conference on
  Knowledge discovery and data mining}  547--556.

\bibitem[Lu and Yang(2012)]{Minggen-JDS-2012}
Lu, M. and Yang, W. (2012).
\newblock Multivariate logistic regression analysis of complex survey data with
  application to {BRFSS} data.
\newblock \emph{Journal of Data Science} \textbf{10}, 157--173.

\bibitem[Meier \emph{et~al.}(2008)Meier, Geer, and Buhlmann]{Meier-JRSSB-2008}
Meier, L., Geer, S. V.~D., and Buhlmann, P. (2008).
\newblock The group lasso for logistic regression.
\newblock \emph{Journal of the Royal Statistical Society: Series B
  (Methodological)} \textbf{70}, 53--71.

\bibitem[Pan \emph{et~al.}(2022)Pan, Chang, Zhu, and Wang]{Pan-SII-2022}
Pan, R., Chang, X., Zhu, X., and Wang, H. (2022).
\newblock Link prediction via latent space logistic regression model.
\newblock \emph{Statistics and Its Interface} \textbf{15}, 267--282.

\bibitem[Pyke and Sheridan(1993)]{Sandra-1993}
Pyke, S. and Sheridan, P. (1993).
\newblock Logistic regression analysis of graduate student retention.
\newblock \emph{Canadian Journal of Higher Education} \textbf{23}, 44--64.

\bibitem[Robin \emph{et~al.}(2011)Robin, Turck, Hainard, Tiberti, Lisacek,
  Sanchez, and Muller]{pROC-2011}
Robin, X., Turck, N., Hainard, A., Tiberti, N., Lisacek, F., Sanchez, J.-C.,
  and Muller, M. (2011).
\newblock proc: an open-source package for r and s+ to analyze and compare roc
  curves.
\newblock \emph{BMC Bioinformatics} \textbf{12:77}.

\bibitem[Schein and Ungar(2007)]{Andrew-2007}
Schein, A. and Ungar, L. (2007).
\newblock Active learning for logistic regression: an evaluation.
\newblock \emph{Machine Learning} \textbf{68}, 235--265.

\bibitem[Shi \emph{et~al.}(2010)Shi, Yin, Osher, and Sajda]{Shi-JMLR-2010}
Shi, J., Yin, W., Osher, S., and Sajda, P. (2010).
\newblock A fast hybrid algorithm for large-scale l1-regularized logistic
  regression.
\newblock \emph{Journal of Machine Learning Research} \textbf{11}, 713--741.

\bibitem[Singh \emph{et~al.}(2009)Singh, Kubica, Larsen, and
  Sorokina]{SIAM-C-L-2009}
Singh, S., Kubica, J., Larsen, S., and Sorokina, D. (2009).
\newblock Parallel large scale feature selection for logistic regression.
\newblock \emph{Proceedings of the 2009 SIAM International Conference on Data
  Mining (SDM)}  1172--1183.

\bibitem[Stefanski and Carroll(1985)]{Stefanski-AOS-1985}
Stefanski, L.~A. and Carroll, R.~J. (1985).
\newblock Covariate measurement error in logistic regression.
\newblock \emph{Annals of Statistics} \textbf{13}, 1335--1351.

\bibitem[Su \emph{et~al.}(2020)Su, Lu, Song, and Huang]{su-JASA-2020}
Su, L., Lu, W., Song, R., and Huang, D. (2020).
\newblock Testing and estimation of social network dependence with time to
  event data.
\newblock \emph{Journal of the American Statistical Association} \textbf{115},
  570--582.

\bibitem[Tripathi \emph{et~al.}(2017)Tripathi, Mahto, and Dey]{ADS-Yogesh-2017}
Tripathi, Y., Mahto, A., and Dey, S. (2017).
\newblock Efficient estimation of the {PDF} and the {CDF} of a generalized
  logistic distribution.
\newblock \emph{Annals of Data Science} \textbf{4}, 63--81.

\bibitem[Wang(2020)]{wang-ICML-2020}
Wang, H. (2020).
\newblock Logistic regression for massive data with rare events.
\newblock \emph{Proceedings of the 37th International Conference on Machine
  Learning} \textbf{119}, 9829--9836.

\bibitem[Wang \emph{et~al.}(2018)Wang, Zhu, and Ma]{wang2018optimal}
Wang, H., Zhu, R., and Ma, P. (2018).
\newblock Optimal subsampling for large sample logistic regression.
\newblock \emph{Journal of the American Statistical Association} \textbf{113},
  522, 829--844.

\bibitem[Wu \emph{et~al.}(2009)Wu, Chen, Hastie, Sobel, and Lange]{WuTT-2009}
Wu, T.~T., Chen, Y.~F., Hastie, T., Sobel, E., and Lange, K. (2009).
\newblock Genome-wide association analysis by lasso penalized logistic
  regression.
\newblock \emph{Bioinformatics} \textbf{25}, 714--721.

\bibitem[Yan \emph{et~al.}(2019)Yan, Jiang, Fienberg, and Leng]{YanT-JASA-2019}
Yan, T., Jiang, B., Fienberg, S., and Leng, C. (2019).
\newblock Statistical inference in a directed network model with covariates.
\newblock \emph{Journal of the American Statistical Association} \textbf{114},
  857--868.

\bibitem[Yang and Loog(2018)]{Pattern-2018}
Yang, Y. and Loog, M. (2018).
\newblock A benchmark and comparison of active learning for logistic
  regression.
\newblock \emph{Pattern Recognition} \textbf{83}, 401--415.

\bibitem[Yuan \emph{et~al.}(2012)Yuan, Ho, and Lin]{Yuan-JMLR-2012}
Yuan, G., Ho, C., and Lin, C. (2012).
\newblock An improved glmnet for {L}1-regularized logistic regression.
\newblock \emph{Journal of Machine Learning Research} \textbf{13}, 1999--2030.

\bibitem[Zhang \emph{et~al.}(2022)Zhang, Guo, and Chang]{Zhang-JCGS-2022}
Zhang, H., Guo, X., and Chang, X. (2022).
\newblock Randomized spectral clustering in large-scale stochastic block
  models.
\newblock \emph{Journal of Computational and Graphical Statistics}  DOI:
  10.1080/10618600.2022.2034636.

\bibitem[Zhao \emph{et~al.}(2022)Zhao, Liu, Wang, and Leng]{Zhao-Biom-2022}
Zhao, J., Liu, X., Wang, H., and Leng, C. (2022).
\newblock Dimension reduction for covariates in network data.
\newblock \emph{Biometrika} \textbf{109}, 85--102.

\bibitem[Zhu \emph{et~al.}(2021)Zhu, Cai, and Ma]{Zhu-JASA-2021}
Zhu, X., Cai, Z., and Ma, Y. (2021).
\newblock Network functional varying coefficient model.
\newblock \emph{Journal of the American Statistical Association}  DOI:
  10.1080/01621459.2021.1901718.

\bibitem[Zhu \emph{et~al.}(2017)Zhu, Pan, Li, Liu, and Wang]{Zhu-AOS-2017}
Zhu, X., Pan, R., Li, G., Liu, Y., and Wang, H. (2017).
\newblock Network vector autoregression.
\newblock \emph{Annals of Statistics} \textbf{43}, 1096--1123.

\bibitem[Zuo \emph{et~al.}(2021)Zuo, Zhang, Wang, and Sun]{zuo2021optimal}
Zuo, L., Zhang, H., Wang, H., and Sun, L. (2021).
\newblock Optimal subsample selection for massive logistic regression with
  distributed data.
\newblock \emph{Computational Statistics} \textbf{36}, 2535--2562.

\end{thebibliography}
\clearpage


\end{document}